\documentclass[sigconf,screen]{acmart}
\usepackage{color,xcolor}
\usepackage{multirow}
\usepackage{booktabs} 
\usepackage{tablefootnote}
\usepackage{ifthen}
\usepackage{url}
\usepackage{amsmath}
\usepackage{threeparttable}
\usepackage{hyphenat}
\usepackage{bbding}
\usepackage{xspace}
\usepackage[T1]{fontenc}
\usepackage[ruled,linesnumbered]{algorithm2e}
\usepackage{array,multirow,graphicx}
\usepackage{float}
\usepackage{balance}
\usepackage{hyperref}
\usepackage{tikz}
\usepackage{calc}
\usepackage{subfigure}
\usepackage{listings,amsfonts}
\usepackage{amsmath,pifont}
\usepackage{url}
\usepackage{balance}
\usepackage{xspace}
\usepackage{hyperref,epsfig,endnotes}
\usepackage{array}
\usepackage{paralist}
\usepackage{multirow,makecell}
\usepackage[normalem]{ulem}
\usepackage{pgfplots}
\usepackage{tcolorbox}
\usepackage{courier}
\usepackage{hyphenat}

\usepackage[skip=0.5pt,font=small,labelfont=bf]{caption}

\def\checkmark{\tikz\fill[scale=0.4](0,.35) -- (.25,0) -- (1,.7) -- (.25,.15) -- cycle;}

\pgfplotsset{width=7cm,compat=1.16}
\usepgfplotslibrary{statistics}
\graphicspath{ {./} }

\definecolor{dkgreen}{rgb}{0,0.6,0}
\definecolor{gray}{rgb}{0.4,0.4,0.4}
\definecolor{mauve}{rgb}{0.58,0,0.82}
\definecolor{darkblue}{rgb}{0.0,0.0,0.6}
\definecolor{lightblue}{rgb}{0.0,0.0,0.9}
\definecolor{cyan}{rgb}{0.0,0.6,0.6}
\definecolor{darkred}{rgb}{0.6,0.0,0.0}

\definecolor{yellow}{RGB}{255,255,153}
\definecolor{grey}{RGB}{220,220,220}
\definecolor{green}{RGB}{0,100,0}

\definecolor{KWColor}{rgb}{0.37,0.08,0.25}
\definecolor{CommentColor}{rgb}{0.133,0.545,0.133}
\definecolor{StringColor}{rgb}{0,0.126,0.941}
\definecolor{commentgreen}{RGB}{2,112,10}
\definecolor{eminence}{RGB}{108,48,130}
\definecolor{weborange}{RGB}{255,165,0}
\definecolor{frenchplum}{RGB}{129,20,83}

\lstdefinelanguage{XML}
{
  morestring=[s][\color{mauve}]{"}{"},
  morestring=[s][\color{black}]{>}{<},
  morecomment=[s]{<?}{?>},
  morecomment=[s][\color{dkgreen}]{<!--}{-->},
  stringstyle=\color{black},
  identifierstyle=\color{lightblue},
  keywordstyle=\color{red},
  morekeywords={xmlns,xsi,noNamespaceSchemaLocation,type,id,x,y,source,target,version,tool,transRef,roleRef,objective,eventually},
  texcl=true
}

\definecolor{yellow}{RGB}{255,255,153}
\definecolor{grey}{RGB}{224,224,224}
\definecolor{green}{RGB}{0,100,0}

\newboolean{showcomments}
\setboolean{showcomments}{true}
\ifthenelse{\boolean{showcomments}}
 { \newcommand{\mynote}[2]{
      \fbox{\bfseries\sffamily\scriptsize#1}
        {\small$\blacktriangleright$\textsf{\emph{#2}}$\blacktriangleleft$}}}
        { \newcommand{\mynote}[2]{}}

\definecolor{DarkOrange}{rgb}{0.8,0.3,0.0} 
\definecolor{DarkCyan}{rgb}{0.0, 0.55, 0.55}
\definecolor{codegreen}{rgb}{0,0.6,0}
\definecolor{codegray}{rgb}{0.5,0.5,0.5}
\definecolor{codepurple}{rgb}{0.58,0,0.82}
\definecolor{backcolour}{rgb}{0.95,0.95,0.92}

\newcommand\tool[1]{\emph{AndroMevol}}

\setcopyright{acmcopyright}
\acmPrice{15.00}
\acmDOI{10.1145/3533767.3534407}
\acmYear{2022}
\copyrightyear{2022}
\acmSubmissionID{issta22main-p486-p}
\acmISBN{978-1-4503-9379-9/22/07}
\acmConference[ISSTA '22]{Proceedings of the 31st ACM SIGSOFT International Symposium on Software Testing and Analysis}{July 18--22, 2022}{Virtual, South Korea}
\acmBooktitle{Proceedings of the 31st ACM SIGSOFT International Symposium on Software Testing and Analysis (ISSTA '22), July 18--22, 2022, Virtual, South Korea}

\begin{document}



\title{Automatically Detecting API-induced Compatibility Issues in Android Apps: A Comparative Analysis (Replicability Study)}

\author{Pei Liu}
\email{Pei.Liu@monash.edu}
\affiliation{
  \institution{Monash University}
  \country{Australia}}

\author{Yanjie Zhao}
\email{Yanjie.Zhao@monash.edu}
\authornote{The first and second authors contributed equally to this research.}
\affiliation{
  \institution{Monash University}
  \country{Australia}}
  
\author{Haipeng Cai}
\email{Haipeng.Cai@wsu.edu}
\affiliation{
  \institution{Washington State University, Pullman}
  \country{United States}}

\author{Mattia Fazzini}
\email{mfazzini@umn.edu}
\affiliation{
  \institution{University of Minnesota}
  \country{United States}}

\author{John Grundy}
\email{John.Grundy@monash.edu}
\affiliation{
  \institution{Monash University}
  \country{Australia}}

\author{Li Li}
\email{Li.Li@monash.edu}
\authornote{Corresponding author.}
\affiliation{
  \institution{Monash University}
  \country{Australia}}

\begin{abstract}
Fragmentation is a serious problem in the Android ecosystem. This problem is mainly caused by the fast evolution of the system itself and the various customizations independently maintained by different smartphone manufacturers.
Many efforts have attempted to mitigate its impact via approaches to automatically pinpoint compatibility issues in Android apps.
Unfortunately, at this stage, it is still unknown if this objective has been fulfilled, and the existing approaches can indeed be replicated and reliably leveraged to pinpoint compatibility issues in the wild.
We, therefore, propose to fill this gap by first conducting a literature review within this topic to identify all the available approaches.
Among the nine identified approaches, we then try our best to reproduce them based on their original datasets.
After that, we go one step further to empirically compare those approaches against common datasets with real-world apps containing compatibility issues.
Experimental results show that existing tools can indeed be reproduced, but their capabilities are quite distinct, as confirmed by the fact that there is only a small overlap of the results reported by the selected tools.
This evidence suggests that more efforts should be spent by our community to achieve sound compatibility issues detection.
\end{abstract}



\maketitle

\section{Introduction}
\label{sec:introduction}


Fragmentation has been a severe problem for the Android ecosystem for years. This refers to the fact that there are a massive number of Android devices manufactured by different companies running different Android operating system versions, including both official and customized ones.
This introduces inconsistencies in that certain apps can only function properly on devices running certain Android versions with certain device features (i.e., the apps crash on other devices), leading to so-called compatibility issues.

Compatibility issues have been considered 
one of the most severe problems in the Android ecosystem. On the one hand, 
they negatively 
impact the users' experience, as apps with compatibility issues may not be able to install on users' devices or may crash at runtime even if successfully installed. This results in poor user experience not only for the app per se but also the whole Android ecosystem. On the other hand, they also increase the difficulties of developing apps.
The vast number of device-Android version combinations create many technical complexities for developers and testers, who must take into account a dizzying number of devices and OS versions, which are non-trivial and yet expensive to achieve without a proper infrastructure in place.

To address these issues, there has been a great deal of research in analyzing the compatibility issues of Android apps.
In the area of static analysis, researchers have proposed various automated approaches to pinpoint one of the most common compatibility issues: API-induced compatibility issues.
For example, Li et al.~\cite{li2018cid} have designed and implemented a prototype tool called CiD that mines the evolution of the official Android framework codebase to locate evolution-induced incompatible Android APIs, i.e., new methods introduced in or existing methods being removed from the latest framework versions.
Wei et al.~\cite{wei2019pivot} have proposed a prototype tool called Pivot for characterizing device-specific incompatible APIs, e.g., APIs that are available for certain devices but not for others.

However, it is still unclear what the status quo of Android app compatibility analyses is, what their strengths and weaknesses are, and to what extent they are able to identify all the possible incompatible Android APIs and their induced compatibility issues in real-world Android apps.
Furthermore, it is also unknown to what extent can we reproduce their experimental results and how well do the tools compare with each other in terms of detecting compatibility issues.
Specifically, in this work, we formulate these concerns into three research questions that we aim to answer through empirical evidence and experimental results.
The three research questions are summarized as follows.

\begin{itemize}
\item \textbf{RQ1: What is the status quo of Android compatibility issues detection approaches?}

We propose to answer this research question through a systematic literature review, aiming to identify the primary studies relevant to statically detecting Android app compatibility issues.

Our review identified nine primary publications that have proposed automated approaches to characterizing Android app compatibility issues.
After careful analysis we summarize five identified types of API-induced compatibility issues: Evolution-induced (Method), Evolution-induced (Field), Device-specific (Method), Device-specific (Field), and  Override/Callback.
Unfortunately none of the existing  approaches can tackle all five types of API-induced compatibility issues.
The most recent, ACID~\cite{mahmud2021android}, can only handle three out of the aforementioned five types.

\item \textbf{RQ2: Can we replicate the experimental results yielded by state-of-the-art tools targeting compatibility issue detection?}

Replicability study has been regarded as an essential method to confirm the reliability of existing research (including both experiments and datasets) and hence has been considered an important field in the software engineering community. 
In the second research question, we aim to confirm the reliability of existing compatibility issues detection approaches by reproducing their experimental results against their original datasets.

Our experimental results show that the majority of experimental results could indeed be reproduced.
The remaining small number of inconsistent results (yielded by IctApiFinder and FicFinder) are mainly caused by unnecessary updates of the tools (such as dependency fixes) and apps (due to unrecorded Github version of the apps).

\item \textbf{RQ3: How well do the tools compare with each other?}

To answer this question and to make a fair comparison, we launch the selected tools on two common sets of benchmark apps: (1) 65 apps used by the authors of selected tools and (2) 645 apps selected from the AndroidCompass dataset~\cite{nielebock2021androidcompass}.

Experimental results show that (1) compatibility issues detection approaches that achieve their purpose via systematically harvested incompatible API rules (such as CiD and IctApiFinder) can identify significantly more issues than those having their rules summarized manually, and (2) the intersection among the results reported by the selected tools is relatively small.

\end{itemize}

\textbf{Open source.} 
The source code and datasets are all made publicly available in our artifact package via the following link:
\begin{center}
    \url{https://zenodo.org/record/6516441}
\end{center}


\section{Status Quo Understanding (RQ1)}

Towards checking how far we are in
automating compatibility issues detection in Android apps, we performed a systematic literature review to understand the status quo about Android app compatibility analyses.

\subsection{Literature Review}

\begin{figure}[!h]
    \centering
    \includegraphics[width=\linewidth]{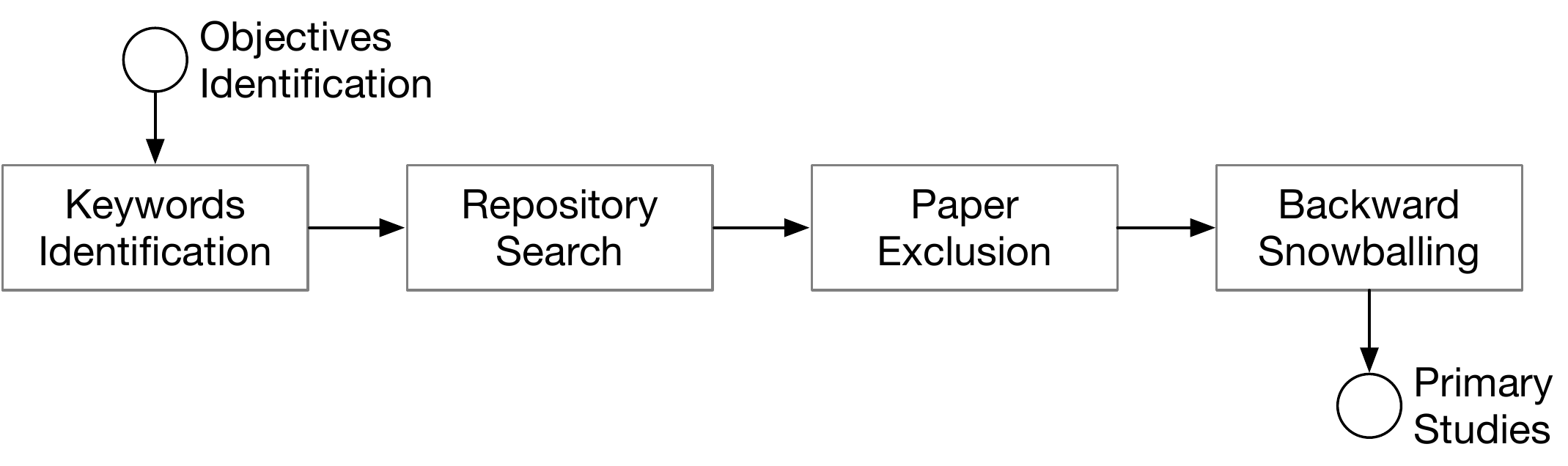}
    \caption{Overview of the literature review process.}
    \label{fig:slr}
\end{figure}

Figure~\ref{fig:slr} illustrates the working processes of the literature review summarized based on the guidelines provided by Keele~\cite{keele2007guidelines} and Brereton et al.~\cite{brereton2007lessons}, as well as lessons learned from our recent practices~\cite{liu2021deep, zhan2021research, shamsujjoha2021developing, kong2018automated}.

\textbf{Keywords Identification.}
To understand the status quo of incompatible app analyses, we resort to a set of keywords to search for relevant publications in popular repositories. The keywords we leveraged are essentially made up of two groups (i.e., G1 and G2). 
Each group contains several keywords.
The search string is then formed as a combination, i.e., g1 AND g2, where g1 and g2 are formed each as a disjunction of the keywords respectively from groups G1 and G2. 

$G1: android, mobile, *phone*$

$G2: *compati*, deprecat*, issue*, evolution$

\textbf{Repository Search.}
To focus the search, we applied these keywords on all the CORE\footnote{\url{https://www.core.edu.au/home}} A/A* ranked venues. This keeps the review process lightweight while ensuring that important related works are not missed.
In the software engineering field (i.e., containing `software' keyword in the venue title and falling in the following fields of research code: 0803 for journals and 4612 for conferences), as summarized in Table~\ref{tab:venues}, there are 19 venues (5 journals and 14 conferences) ranked as A/A* by CORE.
We then go through these 19 venues one by one and apply the aforementioned keywords to search for relevant publications.
Eventually, we were able to locate 44 publications across 13 venues (i.e., there is no relevant paper identified in 6 of the venues).

\begin{table}[ht!]
\centering
\caption{CORE A/A* ranked software engineering venues.}
\resizebox{\linewidth}{!}{
\begin{tabular}{r c p{0.6\linewidth}}
\hline
\textbf{Type} & \textbf{Source} & \textbf{Venues} \\
\hline
Journals & CORE2020 & TOSEM, TSE, EMSE, JSS, IST \\
Conferences & CORE2021 & ASE, ESEC/FSE, ICSE, EASE, ECSA, ISSRE, ESEM, ICSME, MSR, ICSA, SANER, SEAMS, ICST, ISSTA \\
\hline
\end{tabular}}
\label{tab:venues}
\end{table}

\begin{figure}
    \centering
    \includegraphics[width=\linewidth]{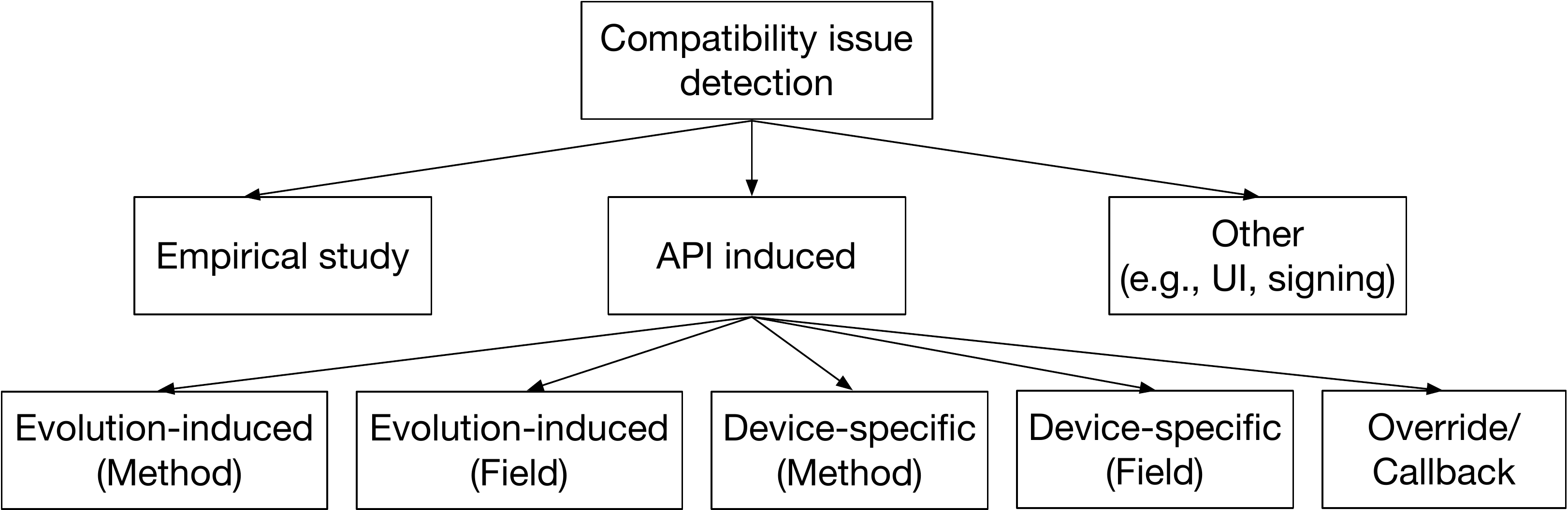}
    \caption{The category of the papers targeting compatibility issues on Android platform.}
    \label{fig:category}
\end{figure}

\textbf{Paper Exclusion.}
As we aimed at collecting as many relevant papers as possible, we have simply considered all the returned results.
However, not every paper is related to automated Android app compatibility issue detection. We there go one step further to read the abstract (and full content if needed) of the obtained papers to only retain the closely related ones by applying the following exclusion criteria: 
(1) Short papers (i.e., less than six pages in double-column format or 11 pages in single-column format) are excluded.
(2) Papers targeting non-Android mobile devices are excluded.
(3) Papers targeting Android but that do not concern compatibility issues are excluded.
(4) Papers targeting Android compatibility issues but that do not concern API-induced ones are excluded (categorized as Other in Figure~\ref{fig:category}).
For example, the work presented by Ki et al.~\cite{ki2019mimic}, which proposes an automated testing framework for Android apps named Mimic for characterizing UI compatibility issues, is excluded. 
Another work presented by Wang et al.~\cite{wang2019characterizing}, which has discussed a type of app signing compatibility issue introduced by unsupported digest/signature algorithms, is also excluded.
(5) Papers targeting Android compatibility issues but that do not introduce automated approaches to detect or resolve them are excluded.
For example, Nielebock et al.~\cite{nielebock2021androidcompass} contribute an Android compatibility check dataset named AndroidCompass, which comprises changes to compatibility checks in the version histories of the Android projects.
Cai et al.~\cite{cai2019large} conduct a large-scale study of compatibility issues based on Android apps developed over the past eight years to comprehend the symptoms and root causes.
These papers do not introduce a prototype tool to detect compatibility issues in Android apps and hence are excluded.
After applying these exclusion criteria, there are 9 papers retained that are closely related to automated incompatible Android API detection.

\textbf{Backward Snowballing.}
Based on the papers identified in the previous steps, we conducted a backward snowballing approach to ensure that important closely related papers (e.g., with titles not matching our search string or published outside of the selected 19 venues) are not missed by our lightweight literature review.
For each paper  we carefully read the related work part and attempted to find cited papers that are closely related to our study but have not yet been included.
This process did not help us identify any new papers, suggesting that the keywords we have selected to search for relevant publications are indeed relevant ones.

\begin{table}[ht!]
\centering
\caption{Full List of Collected and Examined Papers.} 
\label{tab:papers}
\begin{threeparttable}
\resizebox{\linewidth}{!}{
\begin{tabular}{r c c c}
\hline
\textbf{Tool/Reference} &
\textbf{Year} & \textbf{Venue} &
\textbf{Tool availability} \\

\hline
ACID\cite{mahmud2021android} & 2021 & SANER & Available~\cite{websiteacid} \\
\hline
ACRYL (extension)\cite{scalabrino2020api} & 2020 & EMSE & Open Source~\cite{websiteacryl} \\
\hline
ACRYL\cite{scalabrino2019data} & 2019 & MSR & Open Source~\cite{websiteacryl} \\
\hline
Pivot\cite{wei2019pivot} & 2019 & ICSE & Available~\cite{websitePivot} \\
\hline
CiD\cite{li2018cid} & 2018 & ISSTA & Open Source~\cite{websiteCiD} \\
\hline
IctApiFinder\cite{he2018understanding} & 2018 & ASE & Open Source~\cite{websiteictapifinder} \\
\hline
CIDER\cite{huang2018understanding} & 2018 & ASE & Available~\cite{websiteacider} \\
\hline
FicFinder (extension)\cite{wei2018understanding} & 2018 & TSE & Available~\cite{websiteFicFinder} \\
\hline
FicFinder\cite{wei2016taming} & 2016 & ASE & Available~\cite{websiteFicFinder} \\
\hline
\end{tabular}}
\end{threeparttable}
\end{table}

\begin{figure}
    \centering
    \includegraphics[width=0.8\linewidth]{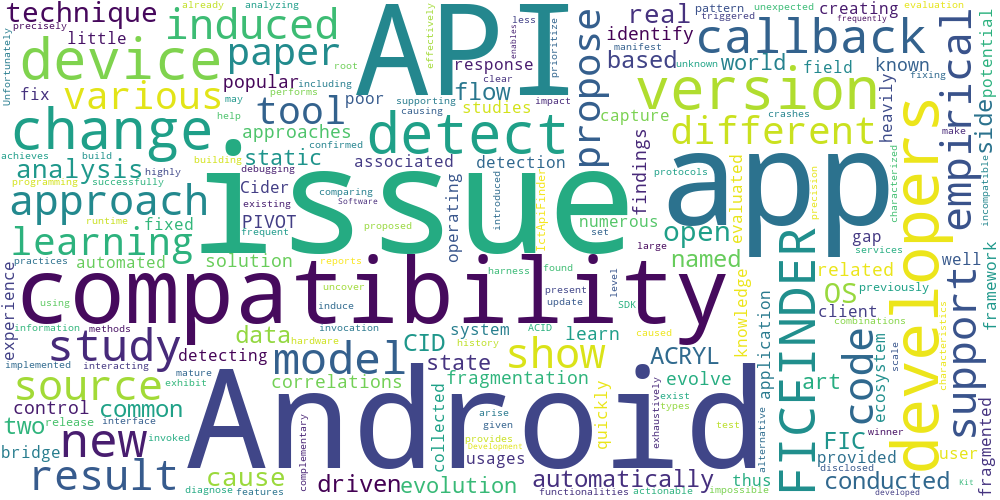}
    \caption{The word cloud of the abstract text in the selected papers.}
    \label{fig:wordcloud}
\end{figure}

\subsection{Result}
\label{subsec:observation}

\begin{figure*}
    \centering
    \includegraphics[width=0.85\linewidth]{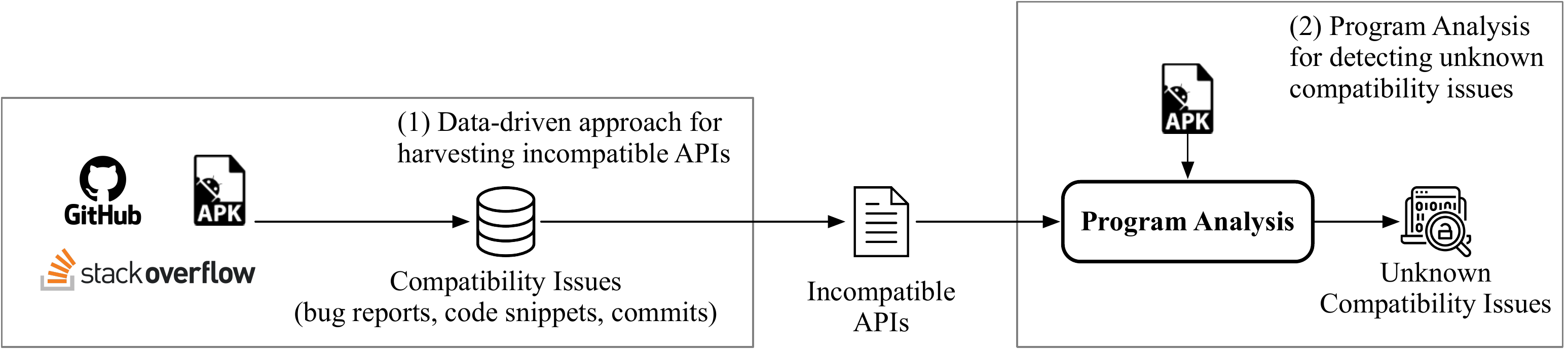}
    \caption{The typical working process of detecting Android compatibility issues.}
    \label{fig:process}
\end{figure*}

In total, our Systematic Literature Review (SLR) search process identified nine relevant papers (hereinafter referred to as primary studies, 
which are listed in the first column of the Table~\ref{tab:papers}.
The nine papers are collected from seven venues with publication dates ranging from 2016 to 2021 (cf. second and third columns in Table~\ref{tab:papers}). The last column describes the availability of these tools. Some of them are open-sourced, while some of them are published as executable files on the associated papers' websites.
Figure~\ref{fig:wordcloud} further illustrates the word cloud of the abstract texts among the identified primary publications.
Terms such as \emph{Android}, \emph{API}, \emph{compatibility}, \emph{issue}, and \emph{app} remain the most representative ones in the word cloud, suggesting that the collected primary publications are indeed relevant to the topic targeted by this work (hence suitable for our study).

\subsection{Status Quo Analysis}

After identifying the primary publications, 
we carefully read their full papers to understand how each of their automated compatibility issues detection approaches are implemented.
We then summarize the common working process taken by those approaches to detect Android compatibility issues.

As shown in Figure~\ref{fig:process}, the objective is often achieved via two steps: (1) data-driven approach for harvesting incompatible APIs and (2) program analysis for detecting unknown compatibility issues.
The output of the first step will be a list of incompatible APIs, which will be taken as input to the second step.
With the two typical steps of working process of compatibility issue detection, we summarized the collected tools as in Table~\ref{tab:apiissue}. The second and third columns describe incompatible APIs collection and issue detection per se separately. CIDER and FicFinder only support issue detection while Pivot only focuses on incompatible APIs harvesting. The remaining tools are working as a whole supporting both APIs harvesting and issue detection.

\begin{table}[ht!]
\centering
\caption{Working Process Support of Tools.}
\label{tab:apiissue}
\begin{threeparttable}
\resizebox{0.9\linewidth}{!}{
\begin{tabular}{r c c}
\hline
\textbf{Tool/Reference} &
\textbf{API Harvest} &
\textbf{Issue Detection} \\
\hline
ACID\cite{mahmud2021android} & \checkmark & \checkmark \\
\hline
ACRYL (extension)\cite{scalabrino2020api} & \checkmark & \checkmark \\
\hline
ACRYL\cite{scalabrino2019data} & \checkmark & \checkmark \\
\hline
Pivot\cite{wei2019pivot} & \checkmark & \XSolidBrush \\
\hline
CiD\cite{li2018cid} & \checkmark & \checkmark \\
\hline
IctApiFinder\cite{he2018understanding} & \checkmark & \checkmark \\
\hline
CIDER\cite{huang2018understanding} & \XSolidBrush & \checkmark \\
\hline
FicFinder (extension)\cite{wei2018understanding} & \XSolidBrush & \checkmark \\
\hline
FicFinder\cite{wei2016taming} & \XSolidBrush & \checkmark \\
\hline
\end{tabular}}
\end{threeparttable}
\end{table}

Among the nine primary publications, as shown in Figure~\ref{fig:category}, after carefully reading their full content, 
we categorize their compatibility issue detection capabilities into five types of issues.
For each of the considered tools, we further summarize and list its capabilities in Table~\ref{tab:detection}. 
Columns 2-6 describe the detection of the five different types of compatibility issues in order as described in Figure~\ref{fig:category}, which are further detailed with concrete examples as follows.

\begin{table*}[ht!]
\centering
\caption{Examination results of the approaches proposed in the retained primary studies.}
\label{tab:detection}
\begin{threeparttable}
\resizebox{0.9\linewidth}{!}{
\begin{tabular}{r | c c c c c | c c}
\hline
\textbf{Tool/Reference} &
\textbf{Evolution-induced} & \textbf{Evolution-induced} &\textbf{Device-specific} &
\textbf{Device-specific} &\textbf{Override/} & \textbf{Systematic} & \textbf{Fully automatic\tnote{2}} \\
&\textbf{(Method)}&\textbf{(Field)}&\textbf{(Method)}&\textbf{(Field)}&\textbf{Callback} & \textbf{(Sound)}  &  \\

\hline
ACID\cite{mahmud2021android} &  \checkmark & \checkmark\tnote{1} & \XSolidBrush & \XSolidBrush & \checkmark &  \checkmark & \checkmark \\
\hline
ACRYL (extension)\cite{scalabrino2020api} &\checkmark & \XSolidBrush & \XSolidBrush & \XSolidBrush& \XSolidBrush & \XSolidBrush & \checkmark \\
\hline
ACRYL\cite{scalabrino2019data} & \checkmark & \XSolidBrush & \XSolidBrush & \XSolidBrush & \XSolidBrush & \XSolidBrush & \checkmark \\
\hline
Pivot\cite{wei2019pivot} & \checkmark & \XSolidBrush & \checkmark & \XSolidBrush & \XSolidBrush & \XSolidBrush & \checkmark \\
\hline
CiD\cite{li2018cid} & \checkmark & \XSolidBrush & \XSolidBrush & \XSolidBrush & \XSolidBrush &  \checkmark & \checkmark \\
\hline
IctApiFinder\cite{he2018understanding} & \checkmark & \checkmark\tnote{1} &\XSolidBrush & \XSolidBrush & \XSolidBrush &  \checkmark & \checkmark \\
\hline
CIDER\cite{huang2018understanding} & \XSolidBrush & \XSolidBrush & \XSolidBrush & \XSolidBrush & \checkmark & \XSolidBrush & \XSolidBrush \\
\hline
FicFinder (extension)\cite{wei2018understanding} & \checkmark & \XSolidBrush & \checkmark & \XSolidBrush & \XSolidBrush & \XSolidBrush & \XSolidBrush\\
\hline
FicFinder\cite{wei2016taming} & \checkmark&\XSolidBrush & \checkmark & \XSolidBrush & \XSolidBrush & \XSolidBrush & \XSolidBrush\\
\hline
\end{tabular}}
\begin{tablenotes}
\footnotesize
\item[1] Only mentioned but not illustrated in detail.
\item[2] There is no human involvement in the core process, e.g., the learning/knowledge collection phase.
\end{tablenotes}
\end{threeparttable}
\end{table*}

\begin{lstlisting}[caption={Code examples.},label={lst:code_example},language=java]
//Example 1: Evolution-induced(Method)
 public class MainActivity extends Activity{
   private TextView mView;
   protected void onCreate(Bundle bundle) { ...
+    if(Build.VERSION.SDK_INT >= 24)
+      wrapper(mView, c, s, null, i);
+    else
       mView.`\textbf{startDrag}`(c, s, null, i);
   }
+  private wrapper(View v, ClipData c, ...) {
+    v.startDragAndDrop(c, s, o, i);
+  }
 }
//Example 2: Evolution-induced(Field)
public static Bitmap getCachedArt(final Context context,final Song song){
  ...
  Options options=new Options();
  options.`\textbf{inDither}`=false;
  options.inPreferredConfig=ARGB_8888;
  ...
}
//Example 3: Device-specific(Method)
  Camera mCamera = Camera.open();
  Camera.Parameters params =  mCamera.getParameters();
  ......
+ if (android.os.Build.MODEL.equals("Nexus 4") {
+  params.`\textbf{setRecordingHint}`(true);
+ }
  ......
  mCamera.setParameters(params);
  mCamera.startPreview();
//Example 4: Device-specific(Field)
private static HttpClient getNewHttpClient() {
  ...
  sf.setHostnameVerifier(`\textbf{SSLSocketFactory.ALLOW\_ALL\_HOSTNAME\_VERIFIER}`);
  ...
}
//Example 5: Override/Callback  
  public void `\textbf{onAttach}`(Context context) {
   super.onAttach(context);
-  mActivity = (BrowserActivity) context;
-  ......
+  attachActivity((BrowserActivity) context);
 }
+ public void onAttach(Activity activity) {
+  super.onAttach(activity);
+  if (Build.VERSION.SDK_INT < 23) {
+    attachActivity((BrowserActivity) activity);
+  }
+ }
+ private void attachActivity(BrowserActivity activity) {
+  mActivity = activity;
+  ......
+ }
\end{lstlisting}

\textbf{Evolution-induced (Method):} The signatures of some public methods are altered (i.e., removed, newly added, or parameter type changes, etc.) during the evolution of the framework.
Example 1 in Listing~\ref{lst:code_example} demonstrates such an example, for which the code snippet is initially reported in \cite{he2018understanding}, where statements beginning with the $+$ signs indicate a possible fix for this incompatibility.
The API \textit{startDrag()} called on Line 8 is introduced into SDK after level 11. However, the minSdkVersion of this app is set to 10. 
Consequently, if not protected with the ``if-else'' block, a ``NoSuchMethodError''  exception will be thrown, leading to crashes on devices running SDK version 10. 

\textbf{Evolution-induced (Field):} During the evolution of the framework, the signatures of some publicly accessible fields could also be altered (i.e., removed or newly added). Unfortunately, apart from \cite{he2018understanding} and \cite{mahmud2021android}, none of the other papers discusses such issues.
Moreover, no relevant examples are given in all the research papers.
Then we use an example that we discovered throughout our research.
There is an evolution-induced issue with a field called "BitmapFactory.Options.inDither" at Line 18 of Example 2 in Listing~\ref{lst:code_example}.
It's supported by API Levels 1 through 23, however since API Level 24, it's been deprecated, creating compatibility issues when an app sets a target SDK version equal to or greater than 24.

\textbf{Device-specific (Method):} 
Due to the customization of smartphone manufacturers, some APIs only work on some devices but not on others.
Example 3 of Listing~\ref{lst:code_example} demonstrates such an example, originally reported by Wei et al.~\cite{wei2019pivot}.
Only if the result of the conditional statement for checking the device identifier according to ``Nexus'' is true, that is, the corresponding app is indeed running on ``Nexus'', the API \emph{setRecordingHint()} on Line 27 will be executed.

\textbf{Device-specific (Field):}
Similar to evolution-induced compatibility issues, the customization of Android frameworks can also introduce incompatible fields (i.e., exist in some devices but not in others), referred to as device-specific fields. 
No code example is provided in our reviewed primary papers, similar to Evolution-induced (Field). We then take the example of "<org.apache.http.conn.ssl.SSL-SocketFactory: org.apache.http.conn.ssl.X509HostnameVerifier ALLOW\_ALL\_HOSTNAME\_VERIFIER>", as shown in Example 4 of Listing~\ref{lst:code_example}.
According to our analysis results, this field is not supported by OPPO smartphones in the SDK of API Level 26, which account for more than 10\% of global smartphone shipments~\cite{OPPOmarket}. 
If an app that uses this field is installed and run on an OPPO smartphone with SDK version 26, compatibility issues may arise.

\textbf{Override/Callback:} Due to the evolution of the Android framework, some callbacks may have been altered. Here, the callbacks are methods defined by the framework that could be explicitly overridden\footnote{Actually, all the methods that are declared as public or protected could eventually be explicitly overridden by client apps. In this work, we take all of such methods into account and hence will not differentiate (and hence specifically emphasize) if the given method is a callback.} by client Android developers, and their execution will be triggered by the framework.
The Example 5 in Listing~\ref{lst:code_example} demonstrates such an example excerpted from \cite{huang2018understanding}. 
The \textit{onAttach(Context)} callback method at Line 39 is introduced from API level 23. This callback method will not be executed if this code is run on a smartphone with an API level lower than 23. Thus it could cause the \textit{mActivity} field not to be initialized, and a ``NullPointerException'' may be thrown when using it.




The table shows obviously that most of the tools are developed for detecting compatibility issues induced by the method evolution of the Android system. For the field evolution-induced compatibility issues, ACID and IctApiFinder have mentioned the issue in the corresponding papers but did not explain the issue in detail. Pivot and FicFinder also considered compatibility issues induced by methods provided by specific devices, while none of the detection tools examined compatibility issues resulted from fields carried by specific devices. For the independent issue induced by the evolution of callback methods, CIDER is the only approach developed intentionally to handle this, while ACID considered both evolution-induced and this special one. 
To summarize, unfortunately, none of these approaches have considered all the identified types of compatibility issues.
The most recent approach, ACID~\cite{mahmud2021android}, can only handle three out of the aforementioned five types, leaving device-specific issues unaddressed.
It is also worth noting that the two approaches, which have indeed taken evolution-based fields into account, have only mentioned this capability but do not elaborate further with the support of experimental evidence.

Furthermore, in column 7 of Table~\ref{tab:detection}, we further summarize whether the proposed approach involves a systematic approach to harvest an incompatible API list (hence the results can be considered complete).
As summarized in Table~\ref{tab:detection}, only three approaches (i.e., ACID, CiD, IctApiFinder) leverage a systematic approach to harvest incompatible APIs.
The majority of considered approaches only take ad-hoc approaches aiming at detecting as many compatibility issues as possible without endeavoring to identify all the possible compatibility issues, i.e., the compatibility issues are not discovered following a systematic approach aiming at covering all the possible cases.
As an example, Scalabrino et al.~\cite{scalabrino2019data} present an automated compatibility issue detection approach called ACRYL, which leverages the knowledge collected from changes implemented in other apps responding to API changes to achieve its purpose.
Such an approach, although implemented in an automated manner, cannot collect all the possible compatibility issues lying in the Android ecosystem and thereby can unfortunately yield false-negative results.

Finally, the last column further highlights whether the proposed approach itself is fully automated or not.
An automated approach should not involve any manual efforts that may pose difficulties to replicate.
Among the selected nine approaches, six of them do provide automated ways to identify compatibility issues (i.e., misuse of incompatible APIs) in real-world Android apps.
Three approaches rely on manual efforts to achieve their objectives, making them not extensible (at least in an easy way) to detect newly introduced compatibility issues.
For example, Wei et al.~\cite{wei2018understanding, wei2016taming} have empirically studied the fragmentation-induced issues to portray the symptoms and root causes of compatibility issues and subsequently proposed a static-analysis tool named FicFinder to detect such compatibility issues.
The major limitation of FicFinder is the requirement of manual efforts to build the patterns of API/context pairs, which are summarized from the aforementioned empirical study.
Such manual efforts are expensive to be extended to summarize more compatibility issues.

\begin{tcolorbox}[title=\textbf{RQ1 Findings}, left=2pt, right=2pt,top=2pt,bottom=2pt]
Our literature review reveals several recent approaches to detecting compatibility issues in Android apps aiming at mitigating the impact of fragmentation in the Android community. Although these state-of-the-art approaches are effective in detecting some issues, they all have weaknesses and limitations.
In our analysis, no state-of-the-art approaches are capable of detecting all five types of compatibility issues that have been identified to date, and many require considerable manual efforts.
There is thus a need for new approaches to holistically resolve all the types of compatibility issues, and these approaches should be systematic and automatic, 
accounting for 
all possible issues in the ecosystem and newly emerging issues.
\end{tcolorbox}

\section{Replicability Study (RQ2)}
\label{sec:rq2}

The second research question aims at checking to what extent can we replicate the experimental results yielded by the state-of-the-art tools targeting compatibility issues detection.

\subsection{Tool Selection}

Ideally, we would like to consider all the tools to perform the replicability study. Among the nine primary studies, there are, in total, seven tools worth reproducing. ACRYL and FicFinder have respectively been first presented in a conference paper and then extended to a journal paper. In these two cases, only the tool versions presented in the latest paper are considered.
Among the seven tools, we decide to exclude Pivot as it does not really involve the actual detection of compatibility issues in Android apps, as highlighted in Table~\ref{tab:apiissue}.
For the remaining six detection tools, we download all of these different tools from their published site and contact the authors of the tools to make sure if the tools \textit{per se} and the experimental datasets are the same as they were presented in the original papers. 
The developers confirm that IctApiFinder~\cite{he2018understanding} has been updated due to the evolution of dependencies. 
We then try to execute them one-by-one in our local environment to make sure they can be successfully reproduced. 
Unfortunately, we have to further exclude ACID and ACRYL from consideration as these two tools cannot be successfully executed.
We have contacted their authors for clarification, but until now, we still cannot properly execute them.
Therefore, we conduct the reproducibility study based on the remaining four tools, which are detailed as follows.

\textbf{CiD~\cite{li2018cid}} first models the lifecycle of Android APIs by extracting Android APIs from Android framework source code and then analyses Android Apps including both the primary app code and extra code. However, it is uncertain whether the Android app has accessed a problematic Android API or not just by checking if the app  contains an invocation of the problematic Android API as the problematic Android API can also be protected by SDK version checkers. Therefore, the authors proposed a path-sensitive inter-procedural backward data-flow analysis to verify if the problematic Android APIs are protected with API-level related conditions. A compatibility issue is identified once the API is not protected by version condition checks and the API is not supported in the range designated in AndroidManifest.xml. 

\textbf{IctApiFinder~\cite{he2018understanding}} first conducts an extensive empirical study over 11 consecutive Android versions and approximately 5,000 Android Apps. The authors find that many different APIs are released between two consecutive Android API releases and thus App developers or third-party library developers provide additional code to guarantee the same behaviors on different OS versions. More importantly, they find that the additional supporting code share the same pattern that is SDK version check. With the provided SDK version check, different Android APIs are invoked to run smoothly on different OS versions. Based on these findings, they propose the tool by first building the inter-procedural control flow graph (ICFG) by Soot for Android Apps and then extracting Android APIs from SDK (android.jar) file as the authors believe that it is not accurate to extract from the SDK document api-version.xml. With the ICFG, it transfers the dataflow analysis problem into a reachability problem. For each Android API in the ICFG, the tool detects if it is supported in the defined API levels interval in AndroidManifest.xml as there are different SDK version constraints (conditional SDK version check to access the Android APIs) in different program points. If the designated API levels are not supported at a certain point, an issue is detected.

\textbf{CIDER~\cite{huang2018understanding}} focuses on compatibility issues caused by callback APIs as the evolution of Android frameworks. With the help of an empirical study, they find that two common types of callback API evolutions: API reachability change and API behavior modification can change app control flows and induce compatibility issues. Thus, they leverage the concept of Callback Control Flow Graph (CCFG)~\cite{yang2015static} and propose a graph-based model, Callback Invocation Protocol Inconsistency Graph (PI-Graph), to capture the structural invocation protocol inconsistencies to detect callback induced compatibility issues (inconsistent app control flows) when apps running on different API levels. The authors first encode seven different PI-Graphs related to 24 key Android APIs from their empirical dataset and then implement the detection tool based on  Soot~\cite{vallee2010soot}.

\textbf{FicFinder~\cite{wei2018understanding}} is actually the first seminal work to better understand fragmentation-induced compatibility issues and detect these issues via the proposed approach. By conducting empirical study and investigating real-world compatibility issues, the authors found that the majority of the issues are induced by the improper use of Android APIs in a problematic running environment, which is called issue-triggering context and the context can be expressed in a context-free grammar. Therefore, the algorithm identifies the issue-inducing Android APIs as well as their dependencies, analyses the calling context, and then compares with the modeled issue-triggering context. To analyze the dependencies issue-inducing API related, the algorithm carries out an inter-procedural backward slicing on callsite to acquire the slices of statements on the basis of program dependence graph~\cite{ferrante1987program}. If the triggering context is not considered before invoking the API, a new issue is reported. To implement this artefact, the well-known static analysis framework, Soot~\cite{vallee2010soot}, is utilized. 
Each of the selected tools requires specific configuration. As the detection result relies on these basic configure parameters, we investigated the tool document and configuration setup process and tried to align the configurations between these selected tools to make sure they do have a similar configuration.

\vspace{-0.5em}
\subsection{Datasets}

Recall that, with RQ2, 
we are interested in evaluating the replicability of the selected tools. We aim to achieve this by running the tools against their original datasets. We therefore request the tools' authors to share their datasets, including mainly the ones with results manually confirmed by the authors\footnote{We decide to not request the full dataset leveraged by the authors because it may involve a very large number of apps that are not convenient to share.} and have been explicitly discussed in their manuscripts (hence can be compared).
To this end, we have eventually selected 65 apps, which are made up of (1) 20 Android apps for CIDER, seven apps for CiD, eight apps for IctApiFinder, 30 for FicFinder.\footnote{The FicFinder authors have actually considered 53 Android projects but only 30 of them can be compiled into Android APKs. Although FicFinder can take either Android APKs or disassembled class files as input, we will only replicate the capability of analyzing Android APKs, which are also the input of the other considered tools.}
It is worth reminding the readers that we have to exclude some of the shared apps because they are no longer available on the web and hence the apps cannot be downloaded based on the information shared by the authors, or the shared source code snippets cannot be compiled to Android apps.
Nevertheless, this exclusion of a small number of apps should not impact the results of the replicability study.

\vspace{-0.5em}
\subsection{Result}

When we do our replication, CIDER and CiD do have exactly the same outputs on the original Android Apps while FicFinder and IctApiFinder have some different outputs regards their original experimental Apps.
We now detail the differences respectively.


\textbf{IctApiFinder.}
The artifact was developed along with the paper in 2018 and was not open-sourced till 2021. With the acquired eight exact Android Apps, we can successfully run the tool on all of them. However, 6 of them do have a different number of issues reported compared with the original paper. Among the six different apps, the paper in total reported 49 issues regardless of TP (True Positive) and FP (False Positive), while our experiment reveals 108 compatibility issues. As we cannot obtain the original results rather than the reported number of issues, we cannot know which issues are different compared with the original results. 
One reason explaining the differences could be that, as also confirmed by the authors, the tool has not been maintained during the last three years. Therefore, there are some dependencies that are not available anymore, and also, there are some APIs not supported in the newer updated dependencies. To release the project, the authors replaced it with newer versions of dependencies and commented out some non-supported APIs in the project.
The authors further noted that they could not make sure if such updates have bad or good effects on the final detection results. 

\textbf{FicFinder.}
The artifact was first published in 2016 and then was extended in 2018. We can successfully execute the artifact on all of the Android projects. The paper describes the detected results in two different categories. The one is compatibility issues in TP and FP, and the other is Good Practice (GP) meaning already fixed issues. After we reproduce in our local environment, seven of them do have different output compared with the original ones presented in the paper. Among the seven apps, we find that 2 of them have the same total number of detected results but have a different number of compatibility issues and good practices, such as GadgeBridge~\cite{Gadgetbridge} was reported one detected issue (regardless of TP and FP) and one GP but we reproduced with 2 issues detected, AnkiDroid~\cite{AnkiDroid} was reported 4 GP detected but we reproduced with 4 issues. The remaining five apps further have a different total number of detected results, such as LibreTorrent~\cite{LibreTorrent} was revealed 6 GP but we detected with only 3 GP, MozStambler~\cite{MozStumbler} contained 1 issue and one GP but we only detected with 1 issue. The possible reason behind this is that they did some regular updates on the artifact as the authors still utilize this one in their research, such as the case study in their newer work Pivot.



To summarize, as revealed by our study, 
most of the experimental results yielded by the selected four tools could be reproduced. The small number of cases that cannot be reproduced are mainly due to tools' updates, either because of lacking maintenance so that we have to arbitrarily update some dependencies to make it runnable in practice or intentional evolutions to keep improving its capabilities.
Such updates, either intended or not, have indeed caused difficulties in reproducing the exact original results. Therefore, we argue that there is a need to always record the artifacts, along with the experimental datasets such as Android apps including both source code and bytecode APK files if possible, in permanent sites (e.g.,  Zenodo or Figshare). The artifacts should also be well-configured in docker-alike containers that can support direct execution of the tools and hence mitigate unnecessary dependency errors that may hinder the tools' replicability.

\begin{tcolorbox}[before skip=0.4cm, after skip=0.6cm, title=\textbf{RQ2 Findings}, left=2pt, right=2pt,top=2pt,bottom=2pt]
Most of the experimental results yielded by the four selected state-of-the-art tools can indeed be reproduced. There are, however, a small number of non-replicated cases that are mainly caused by slightly updates of the tools or the evaluated apps.
\end{tcolorbox}

\section{Comparison Study (RQ3)}

The last research question aims to empirically compare the state-of-the-art tools targeting the detection of compatibility issues in Android apps.
We answer this research question by first presenting the experimental setup (including tool selection and datasets) in Subsections~\ref{subsec:rq3_tools} and~\ref{subsec:rq3_datasets} and then the experimental results in Subsection~\ref{subsec:rq3_results}.

\subsection{Tools Selection}
\label{subsec:rq3_tools}

Recall that there are only four tools that we can replicate to scan compatibility issues (as discussed in the previous section). Therefore, we select the same four tools to achieve this objective in this work, i.e., comparing these four tools w.r.t. their compatibility issues detection capabilities.

\subsection{Datasets}
\label{subsec:rq3_datasets}

In this work, we resort to the following two datasets to support the comparison study.

\begin{itemize}
\item \textbf{Dataset1}: The same 65 apps used for the replicability study as discussed in Section~\ref{sec:rq2}.
    
\item \textbf{Dataset2}: 645 Android apps selected from the AndroidCompass dataset~\cite{nielebock2021androidcompass}.
AndroidCompass contains a dataset of git commits related to Android compatibility checks (including evolution-induced, device-specific, and override/callback-related ones), which are originally harvested from 1,375 open-source Android projects on Github.
Some git commits contain compatibility issue fixes (e.g., adding compatibility checks for APIs that are not protected initially), while others do not (e.g., adding new Java files that include compatibility checks).
In this work, we are only interested in the former ones as based on which we could locate problematic app versions containing actual compatibility issues (i.e., the apps compiled based on their immediate previous commit).
We could further collect the actual compatibility issues based on the compatibility checks added in the fix commits for each of the compiled apps.
This study will leverage this information as partial ground truth to support the comparison study.
Unfortunately, several app projects are no longer available on Github, while some others cannot be easily complied into APKs (e.g., due to missing library dependencies), we have to exclude them.
Eventually, we were able to collect 645 apps to fulfill this dataset.

\end{itemize}


\subsection{Result}
\label{subsec:rq3_results}


\begin{table}[]
    \centering
    \scriptsize
    \caption{Experimental results obtained based on the 65 apps located in Dataset1.}
    \begin{tabular}{|l|c|c|c|c|}
    \hline
        \multirow{2}{6em}{App Name}  & Callback-induced  &
        \multicolumn{3}{c|}{Evolution-Induced} \\
    \cline{2-5}
        & CIDER & FicFinder & IctApiFinder & CiD \\
    \hline
    Tinfoil-Facebook               &  0  &  1  &  1  &  2  \\
    \hline
    kolabnotes                     &  1  &  4  &  5  &  6  \\
    \hline
    SteamGifts-chocolate-debug     &  0  &  6  &  20 &  23 \\
    \hline
    OsmAnd                         &  1  &  3  &  7  &  44 \\
    \hline
    iFixitAndroid                  &  0  &  7  &  58 & 218 \\
    \hline
    Simple-Solitaire               &  1  &  0  &  1  &  10 \\
    \hline
    Anki-Android                   &  0  &  6  & 150 &  86 \\
    \hline
    login-sample-debug             &  4  &  0  &  2  &  3  \\
    \hline
    ooniprobe-android-1.3.1-debug  &  1  &  0  &  4  &  38 \\
    \hline
    APICompatibility\_Inheritance  &  0  &  0  &  2  &  3  \\
    \hline
    APICompatibility\_Varargs      &  0  &  0  &  2  &  2  \\
    \hline
    SurvivalManual-4.1-debug       &  0  &  2  &  1  &  15 \\
    \hline
    Calendula                      &  0  &  15 &  29 &  63 \\
    \hline
    libretorrent                   &  3  &  1  &  13 &  59 \\
    \hline
    APICompatibility\_Protection2  &  0  &  1  &  1  &  0  \\
    \hline
    StreetComplete                 &  0  &  2  &  7  &  5  \\
    \hline
    red-moon                       &  0  &  0  &  13 &  21 \\
    \hline
    padland                        &  1  &  0  &  13 &  4  \\
    \hline
    duckduckgo-0.6.0-release       &  1  &  0  &  1  &  2  \\
    \hline
    transdroid                     &  0  &  1  & 214 &  37 \\
    \hline
    materialistic-hacker-news      &  0  &  1  &  32 &  36 \\
    \hline
    materialfbook                  &  0  &  1  &  15 &  38 \\
    \hline
    ownCloud                       &  0  &  2  &  66 & 181 \\
    \hline
    AndStatus                      &  0  &  2  &  43 &  27 \\
    \hline
    RedReader                      &  0  &  1  &  30 &  7  \\
    \hline
    opentasks-1.1.8.2              &  0  &  24 &  14 &  51 \\
    \hline
    APICompatibility\_Basic        &  0  &  0  &  1  &  1  \\
    \hline
    Gadgetbridge                   &  0  &  2  &  21 &  35 \\
    \hline
    Total                          & 13  & 82  & 766 & 1,017  \\ 
   \hline
    \end{tabular}
    \label{tab:benchresult}
\end{table}

\textbf{Results on Dataset1.}
We first launch the selected tools to analyze the apps in Dataset1.
Unfortunately, 37 apps cannot be handled successfully by both IctApiFinder and CiD (i.e., 24 and 15 failures, respectively).
The corresponding error messages indicate that the failures are mainly raised by Soot, the underlying static analysis framework leveraged by these two tools.
This problem has been discussed by the authors in their article as a potential threat to validity.
It is also a well-known problem when performing static analysis on top of Soot.

For the remaining 28 successfully analyzed apps, Table~\ref{tab:benchresult} presents the detection results.
CIDER, different from the other three detection tools, was developed for callback-induced compatibility issues. 
Among the 28 apps, there are only 8 apps being reported to include callback-induced issues. The reason behind this small number could be explained by the fact that the tool only leverages seven manually summarized rules to detect such issues.
Such a manual process may not be able to include all the different situations and hence may lead to incomplete results.
Similarly, FicFinder, which leverages 20 manually summarized incompatible APIs, reports only 82 compatibility issues, which are also significantly fewer results compared with the remaining two tools that have leveraged systematic approaches to harvest incompatible APIs (as indicated in Table~\ref{tab:apiissue}).
This experimental result further confirms that it is essential to invent systematic approaches to harvest incompatible APIs so as to support automated compatibility issues detection in Android apps.

\begin{figure}
    \centering
    \includegraphics[width=0.6\linewidth]{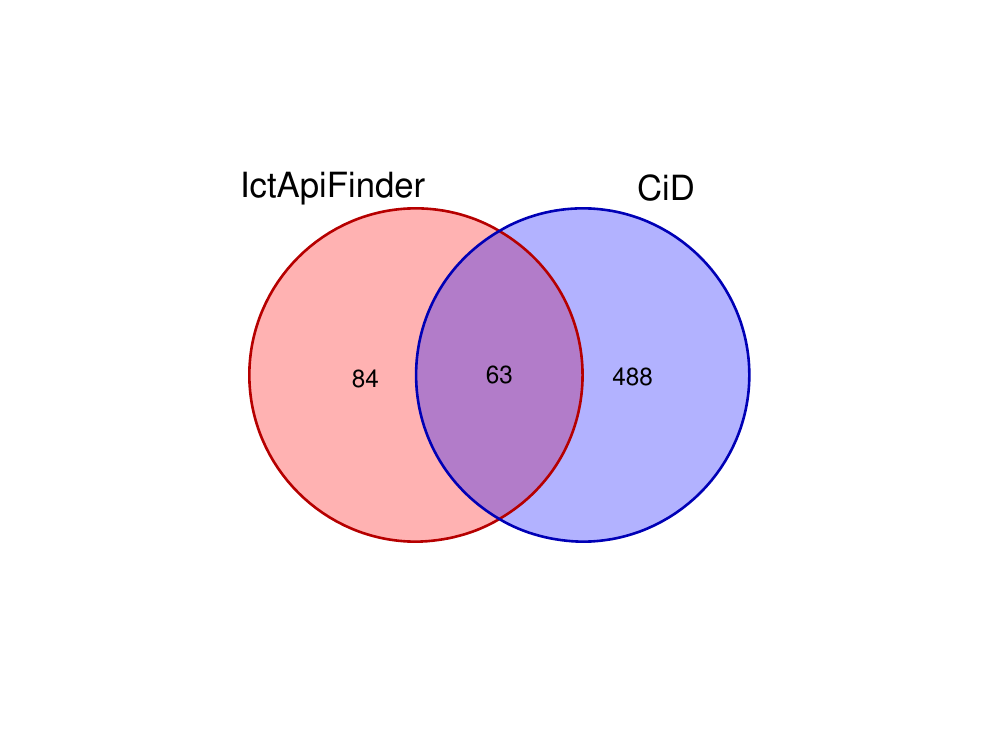}
    \caption{Venn diagram of incompatible APIs utilized in IctApiFinder and CiD.} 
    \label{fig:compvenn}
\end{figure}

While both IctApiFinder and CiD yield significantly more results than FicFinder and they do take systematic approaches to collect incompatible APIs, their results are quite different.
Among the 28 apps successfully analyzed by both of these two tools, IctApiFinder and CiD respectively yields in total 766 and 1,017 issues, for which only 52 reported by both of them.
This experimental result is quite surprising as we would have expected that IctApiFinder and CiD would have much more overlap in terms of their detected compatibility issues.
We therefore go one step deeper to investigate why these two tools yield quite different results, i.e., being able to locate a quite number of compatibility issues while also missing many of them reported by the other tool.
We look at the number of distinct incompatible APIs detected by these two tools.
The 766 and 1,017 compatibility issues reported by IctApiFinder and CiD are essentially caused by 147 and 551 incompatible APIs, respectively.
As highlighted in Figure~\ref{fig:compvenn}, the intersection between these two incompatible APIs sets is quite small (i.e., only 63 out of 551 incompatible APIs considered by CiD are also taken into account by IctApiFinder).
One reason causing this difference is that different time framework of Android framework versions are considered (e.g., the incompatible APIs collected by IctApiFinder are from 4 to 27, while CiD is from API 1 to 25).
Subsequently, the common compatibility issues reported by both of these two tools will be small as well.

\textbf{Results on Dataset2.}
\begin{figure}[!t]
    \centering
    \includegraphics[width=0.85\linewidth]{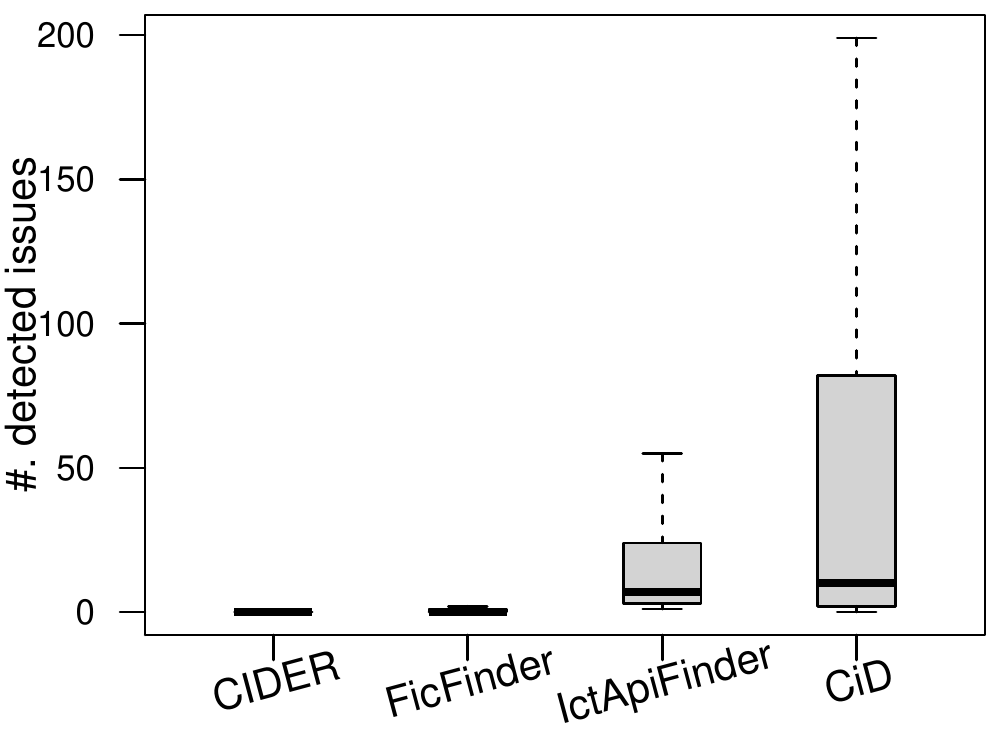}
    \caption{Compatibility issues detected by different detection tools against Dataset2.}
    \label{fig:compdetect}
\end{figure}
We then launch the selected tools on Dataset2, which contains a large number of real-world Android apps selected from the AndroidCompass compatibility checks dataset.
Unfortunately, over half the apps are excluded from the dataset as they cannot be successfully analyzed by all the selected tools.
Among the 277 remaining apps, CIDER, FicFinder, IctApiFinder, and CID have reported 12, 277, 5,009, and 27,874 compatibility issues, respectively.
Figure~\ref{fig:compdetect} further illustrates the distribution of detected compatibility issues in real-world Android Apps.
Clearly, CiD yields more issues than the other tools, followed by IctApiFinder and then FicFinder. CIDER reports the least number of compatibility issues.
These differences are also significant as confirmed by a Mann-Whitney-Wilcoxon (MWW) test at a significant level\footnote{Given a significance level $\alpha$ = 0.001, if p-value < $\alpha$, there is one chance in a thousand that the difference between the datasets is due to a coincidence.} at 0.001.


Observant readers may have noticed that this experiment, although with a large number of apps, supports the same findings discussed previously.
First, there is a strong need to invent systematic approaches to harvest compatibility issues detection rules (i.e., identifying incompatible APIs). As shown in Figure~\ref{fig:compdetect}, the number of issues reported by CIDER and FicFinder (with manually summarized rules) is significantly less than that achieved by IctApiFinder and CiD (with systematically harvested rules).
Furthermore, the fact that the intersection between the results yielded by the selected tools is quite small suggests that existing tools could be leveraged to complement each other.
This result further shows that there is still a gap in the community to implement promising approaches to flag compatibility issues in Android apps, i.e., the capability of detecting compatibility issues has not been mature.
Last but not the least, we believe that it is not exactly fair to directly compare existing tools targeting compatibility issues detection in Android apps as the evolution of the Android ecosystem is very fast.
Tools developed at different times will likely collect a different set of incompatible APIs (e.g., the incompatible APIs collected by IctApiFinder are from 4 to 27, while CiD is from API 1 to 25), which subsequently will lead to a different set of compatibility issues.
Therefore, we argue that, when comparing compatibility issues detection tools, there is a strong need to make sure that the underlying set of incompatible APIs is kept the same, which is however non-trivial to achieve as existing tools may not always be made open-source.



\begin{tcolorbox}[before skip=0.4cm, after skip=0.6cm, title=\textbf{RQ3 Findings}, left=2pt, right=2pt,top=2pt,bottom=2pt]
Comparing the selected four tools on the same datasets, CiD is able to yield more compatibility issues than the other tools, followed by IctApiFinder, and then FicFinder.
Their results are however not well overlapped, suggesting the existing tools are complementary to each other and yet still have limitations to achieve sound compatibility issues detection.
Furthermore, the fact that CiD and IctApiFinder can yield significantly more results than FicFinder and CIDER suggests that it is essential to leverage systematic approaches to mine incompatible APIs so as to support the detection of compatibility issues.
\end{tcolorbox}

\section{Discussion}
\label{sec:discussion}

We discuss the key implications of this research, including prioritized research directions that should be conducted for mitigating the fragmentation impact on the Android community.
Our literature review and experimental findings raise a number of issues and opportunities for research and practice communities.

\vspace{-0.5em}

\subsection{Implication}
\label{subsec:implication}

\textbf{Continuous Improvement to Adapt to the Fast Evolution of the Mobile Ecosystem.}
With the rapid evolution of the open-source Android Operating System, detection tool maintainers need to take the new system releases into account. Besides, many device vendors always release lots of different models as their own publish step. To detect the newly introduced compatibility issues, these tools need to be refined once new version system released and new device induced. However, these tools are not self-adaptive. They all need to be carefully adjusted.

\begin{figure}
    \centering
    \includegraphics[width=0.85\linewidth]{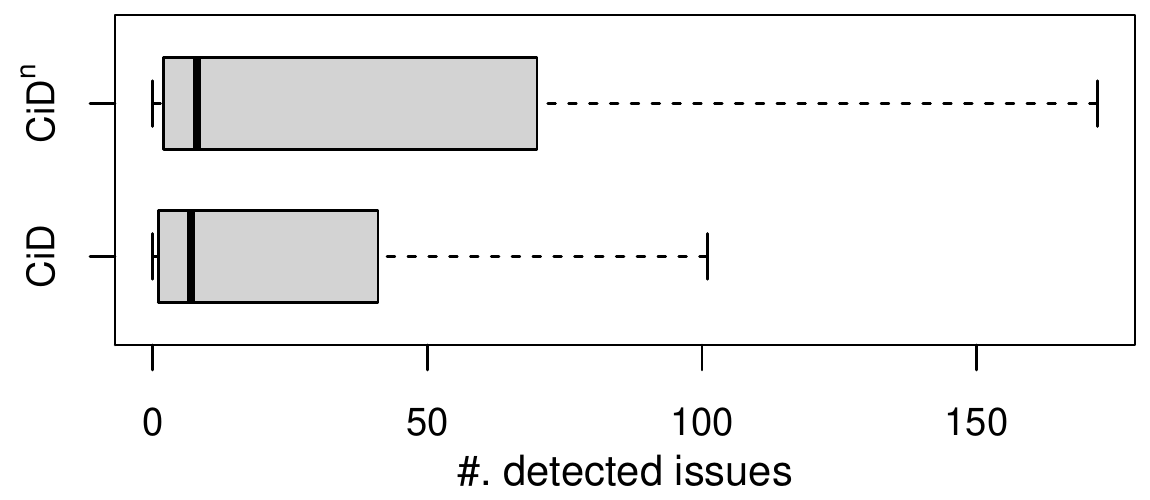}
    \caption{Comparison between original CiD and API life-cycle extended CiD.}
    \label{fig:cidext}
\end{figure}

As an example towards demonstrating the necessity to continuously update the tools to adapt to the fast evolution of the mobile ecosystem, we spend additional efforts to update the open-source CiD project by extending its supported API ranges from 1-25 to 1-31 (Android12 with API level 31 is the latest Android release). The updated version is referred to as CiD$^n$.
We then launch CiD$^n$ to analyze the apps in Dataset2.
Figure~\ref{fig:cidext} summarizes the experimental results, along with that achieved by the original CiD.
Clearly, CiD's performance has indeed been improved after adapting to the latest release of Android frameworks.
This evidence strongly suggests the necessity to keep adapting compatibility issues detection tools to support the latest changes of the mobile ecosystem.
We therefore argue that different automation approaches are needed to facilitate the extraction of Android APIs in order to automate issue detection when new Android versions and devices are released.

\textbf{Extracting Complete Incompatible APIs:}
The state-of-the-art tools all heavily rely on summarized incompatible APIs regardless of whether they are manually summarized or systematically collected. However, tools we analyzed all have their own approach to collect incompatible APIs, which not only sets obstacles to have a fair comparison but also undermines capacities of the detection tools. Therefore, an effective approach to extract a complete list of incompatible APIs is necessary.

\textbf{Integrating Dynamic Testing to Verify Compatibility Issues:}
Currently, most research approaches proposed to tackle compatibility issues in Android apps rely on static analysis.
However, efficient, static analysis is also known to yield many false-positive results.
We argue that dynamic testing approaches should also be included to supplement the analysis of static analysis approaches (e.g., to practically verify the results yielded by static analysis approaches).
It is nevertheless non-trivial to build a comprehensive dynamic testing environment for checking compatibility issues, as it needs to include all publicly available Android devices, for which the number is also continuously changing.
To cope with this, we argue that crowdsourced mobile app testing could be leveraged, especially in lightweight mode directly supported by the Android system, to pinpoint and subsequently mitigate compatibility issues.

\textbf{Characterizing Semantics-changing Incompatible APIs: }
In addition to the five types of incompatible APIs discussed in this work, which are all related to the existence of the APIs, there is another type of API-induced compatibility issue that goes beyond APIs' existence to concern their semantic changes.
Given an API with semantic changes, even if its signature persisted in the framework, the client apps accessed into it could also be impacted. Such semantics changes will be propagated to the client app, which may not have yet adapted to such changes.
As recently revealed by Liu et al.~\cite{liu2021identifying}, there are indeed a number of Android APIs involving semantic changes during the evolution of the framework.
However, such semantic changes are hard to be automatically identified, so as to the corresponding compatibility issues.
Therefore, we argue that our community should also pay special attention to semantics-changed incompatible APIs and invent advanced approaches to mitigate them, either by carefully (1) documenting them if it is unavoidable to change the semantics of existing APIs or (2) testing the client apps to identify and fix such issues before publishing the apps to end-users.


\textbf{Supporting Automated Compatibility Issue Repair: }
Finally, after API compatibility issues are identified, we argue that automated approaches are also needed to help developers to fix them. This is especially true for such apps that have already been released to the public, as users may not even be able to install the apps or face runtime crashes even if the apps can be successfully installed.
Automated repairing approaches could keep users from encountering such unfavorable situations, meanwhile helping app developers fix the issues for better future releases.

\subsection{Threats to validity} 

As it is the chase of most empirical studies, there are threats to validity associated with the results we presented. One threat is the configuration of all our selected detection tools. All selected tools are implemented on top of the Soot static analysis framework, which also requires Android frameworks as an input parameter. However, the version of the Soot and Android framework may be still different because we cannot know the exact versions leveraged in every detection tool. To mitigate this threat, we meticulously align the configurations among them as much as we can to provide an approximately same environment.  
Another threat depends on the approach to harvest incompatible APIs, especially between IctApiFinder and CiD. IctApiFinder extracts APIs from Android framework API levels 4 to 27 based on published artifacts, while CiD acquires from the source code of Android framework API levels 1 to 25 on their own approach. Different ranges of API levels and the trade-offs made while pinpointing incompatible APIs would unavoidably bring in discrepancies, which may result in different performances even on the same dataset.  In the future, we will try to align the extraction approach to achieve a fair comparison.

\section{Related Work}
\label{sec:relatedwork}

In recent years, compatibility issues have been a hot topic in the Android community~\cite{rolim2017learning, lawall2018coccinelle, wang2019characterizing, scalabrino2019data, kang2019semantic, collie2020m3, zhao2022towards}. 
Since the apps are inseparable from the official Android APIs, it is essential to probe compatibility issues caused by the evolution of the Android operating systems.

Besides the tools we investigated in the paper, there are many other works handling various API issues.
For example, Li et al.~\cite{li2018characterising, li2020cda} build a prototype tool, CDA, to characterize deprecated Android APIs by mining the evolution of the Android framework.
Similar method has also been applied to characterize inaccessible APIs~\cite{li2016accessing} and inconsistent release time of Android apps~\cite{li2018moonlightbox}.
Scalabrino et al.~\cite{scalabrino2019data} introduce ACRYL, learning from the change histories of other apps in response to API evolution.
It can identify compatibility issues, yet in addition suggest repairs.
The authors empirically compare ACRYL and CiD and track down no obvious winner, but the results indicate the possibility of combining the two methods in the future. 
Later on, they extend their work~\cite{scalabrino2020api} by enlarging the datasets and adding some interviews and details, but there is no obvious improvement in terms of the detection approach.
Xia et al.~\cite{xia2020android} conduct a large-scale study on the practice of handling OS-induced API compatibility issues and their solutions, and they propose a tool named RAPID to ascertain whether a compatibility issue has been resolved.
Mobilio et al.~\cite{mobilio2020filo} acquaint a tool named FILO which can assist Android designers in tackling backward compatibility issues caused by API upgrades. 
FILO is designed to recognize app methods that need to be altered to adapt to the API changes and report symptoms observed in failed executions to facilitate repair.
Mahmud et al.~\cite{mahmud2021android} propose ACID, an approach to detecting compatibility issues caused by API evolution. 
Experimental results demonstrate that ACID is more accurate and faster in detecting compatibility issues than previous techniques.
The fly in the ointment is that ACID only considers the changes in Android method invocations and callbacks brought about by evolution rather than considering device-specific compatibility issues.

To detect such compatibility issues, different information flows are needed to identify by constructing inter- and intra-procedural control flow graph~\cite{li2017static}. Qiu et al.~\cite{qiu2018analyzing} did an extensive comparison among three most prominent static analysis tools including FlowDroid~\cite{arzt2014flowdroid} combined with IccTA~\cite{li2015iccta}, DroidRA~\cite{li2016droidra,sun2020taming}, AmanDroid~\cite{wei2014amandroid,wei2018amandroid}, and DroidSafe~\cite{gordon2015information}. They spotted out the advantages and shortcomings of each tools and revealed that it is important to provide detailed configuration and setup environment specification to guarantee the replicability of experiments.

In non-Android communities, research on compatibility issues is also pervasive~\cite{robbes2012developers, hora2015developers, zhou2016api, brito2016developers, sawant2018reaction}.
Sawant et al.~\cite{sawant2018reaction} analyze clients of popular third-party Java APIs and the JDK API and publicise a large dataset; also, they look into the connection between the client's response patterns and the deprecation policy the related API adopted.
Chen et al.~\cite{chen2020taming} present an approach named DeBBI, which leverages the test suites of various client projects to detect library behavioral backward incompatibilities.

To compare different tools developed for the same issue, Su et. al.~\cite{su2021benchmarking} did an extensive comparison and proposed a new benchmark called Themis facilitating our research community for automated GUI testing. They collected critical bugs reported on Github with respect to their bug label revealing the severity and did experiments with five state-of-the-art testing tools integrated with Monkey~\cite{Monkey}, and then gave out qualitative and quantitative analysis result. They successfully identified 5 different challenges that these tools still face, such as the reachability of deep use scenarios, test input generation etc., and shed lights on future research based on their systematic analysis results, such as integrating heuristics to improve the capability to spot GUI bugs.

\section{Conclusion} 
\label{sec:conclusion}

In this paper, we have conducted a literature review on research works targeting Android app compatibility issues. 
Based on this review, we are able to identify nine state-of-the-art works proposed to detect compatibility issues in Android apps and among which we have summarized five types of incompatible issues reported by our fellow researchers.
We then confirm the reproducibility of the selected tools based on a replication study by running the tools against their original datasets.
We further go one step deeper to conduct an empirical comparison study among the selected tools.
Our findings indicate that compatibility issues detection is still at an early stage, which requires attention from the community to keep improving so as to achieve sound compatibility issues detection.

\section*{Acknowledgements}

This work is supported by ARC Laureate Fellowship FL190100035, ARC Discovery Early Career Researcher Award (DECRA) project DE200100016, and a Discovery project DP200100020.

\balance
\bibliographystyle{ACM-Reference-Format}
\bibliography{main.bib}


\begin{thebibliography}{63}


\ifx \showCODEN    \undefined \def \showCODEN     #1{\unskip}     \fi
\ifx \showDOI      \undefined \def \showDOI       #1{#1}\fi
\ifx \showISBNx    \undefined \def \showISBNx     #1{\unskip}     \fi
\ifx \showISBNxiii \undefined \def \showISBNxiii  #1{\unskip}     \fi
\ifx \showISSN     \undefined \def \showISSN      #1{\unskip}     \fi
\ifx \showLCCN     \undefined \def \showLCCN      #1{\unskip}     \fi
\ifx \shownote     \undefined \def \shownote      #1{#1}          \fi
\ifx \showarticletitle \undefined \def \showarticletitle #1{#1}   \fi
\ifx \showURL      \undefined \def \showURL       {\relax}        \fi
\providecommand\bibfield[2]{#2}
\providecommand\bibinfo[2]{#2}
\providecommand\natexlab[1]{#1}
\providecommand\showeprint[2][]{arXiv:#2}

\bibitem[web(2021a)]%
        {websiteacid}
 \bibinfo{year}{2021}\natexlab{a}.
\newblock \bibinfo{title}{ACID}.
\newblock \bibinfo{howpublished}{\url{https://github.com/TSUMahmud/acid}}.
\newblock


\bibitem[web(2021b)]%
        {websiteacryl}
 \bibinfo{year}{2021}\natexlab{b}.
\newblock \bibinfo{title}{ACRyL}.
\newblock
  \bibinfo{howpublished}{\url{https://github.com/intersimone999/acryl}}.
\newblock


\bibitem[web(2021c)]%
        {websiteCiD}
 \bibinfo{year}{2021}\natexlab{c}.
\newblock \bibinfo{title}{CiD}.
\newblock \bibinfo{howpublished}{\url{https://github.com/lilicoding/CiD}}.
\newblock


\bibitem[web(2021d)]%
        {websiteacider}
 \bibinfo{year}{2021}\natexlab{d}.
\newblock \bibinfo{title}{CIDER}.
\newblock
  \bibinfo{howpublished}{\url{https://github.com/cideranalyzer/cideranalyzer.github.io}}.
\newblock


\bibitem[web(2021e)]%
        {websitePivot}
 \bibinfo{year}{2021}\natexlab{e}.
\newblock \bibinfo{title}{Download Pivot}.
\newblock \bibinfo{howpublished}{\url{https://ficissuepivot.github.io/Pivot/}}.
\newblock


\bibitem[web(2021f)]%
        {websiteFicFinder}
 \bibinfo{year}{2021}\natexlab{f}.
\newblock \bibinfo{title}{FicFinder Project Homepage}.
\newblock \bibinfo{howpublished}{\url{http://sccpu2.cse.ust.hk/ficfinder/}}.
\newblock


\bibitem[web(2021g)]%
        {websiteictapifinder}
 \bibinfo{year}{2021}\natexlab{g}.
\newblock \bibinfo{title}{IctApiFinder}.
\newblock
  \bibinfo{howpublished}{\url{https://github.com/DongjieHe/IctApiFinder}}.
\newblock


\bibitem[OPP(2021)]%
        {OPPOmarket}
 \bibinfo{year}{2021}\natexlab{}.
\newblock \bibinfo{booktitle}{\emph{OPPO's share of smartphone shipments
  worldwide}}.
\newblock
\urldef\tempurl%
\url{https://www.statista.com/statistics/628545/global-market-share-held-by-oppo-smartphones/}
\showURL{%
\tempurl}


\bibitem[Ank(2022)]%
        {AnkiDroid}
 \bibinfo{year}{2022}\natexlab{}.
\newblock \bibinfo{title}{AnkiDroid}.
\newblock
  \bibinfo{howpublished}{\url{https://github.com/ankidroid/Anki-Android}}.
\newblock


\bibitem[Gad(2022)]%
        {Gadgetbridge}
 \bibinfo{year}{2022}\natexlab{}.
\newblock \bibinfo{title}{Gadgetbridge}.
\newblock
  \bibinfo{howpublished}{\url{https://github.com/Freeyourgadget/Gadgetbridge}}.
\newblock


\bibitem[Lib(2022)]%
        {LibreTorrent}
 \bibinfo{year}{2022}\natexlab{}.
\newblock \bibinfo{title}{LibreTorrent}.
\newblock
  \bibinfo{howpublished}{\url{https://github.com/proninyaroslav/libretorrent}}.
\newblock


\bibitem[Mon(2022)]%
        {Monkey}
 \bibinfo{year}{2022}\natexlab{}.
\newblock \bibinfo{title}{Monkey}.
\newblock
  \bibinfo{howpublished}{\url{http://developer.android.com/tools/help/monkey.html}}.
\newblock


\bibitem[Moz(2022)]%
        {MozStumbler}
 \bibinfo{year}{2022}\natexlab{}.
\newblock \bibinfo{title}{MozStumbler}.
\newblock \bibinfo{howpublished}{\url{https://github.com/mozilla/MozStumbler}}.
\newblock


\bibitem[Arzt et~al\mbox{.}(2014)]%
        {arzt2014flowdroid}
\bibfield{author}{\bibinfo{person}{Steven Arzt}, \bibinfo{person}{Siegfried
  Rasthofer}, \bibinfo{person}{Christian Fritz}, \bibinfo{person}{Eric Bodden},
  \bibinfo{person}{Alexandre Bartel}, \bibinfo{person}{Jacques Klein},
  \bibinfo{person}{Yves Le~Traon}, \bibinfo{person}{Damien Octeau}, {and}
  \bibinfo{person}{Patrick McDaniel}.} \bibinfo{year}{2014}\natexlab{}.
\newblock \showarticletitle{Flowdroid: Precise context, flow, field,
  object-sensitive and lifecycle-aware taint analysis for android apps}.
\newblock \bibinfo{journal}{\emph{Acm Sigplan Notices}} \bibinfo{volume}{49},
  \bibinfo{number}{6} (\bibinfo{year}{2014}), \bibinfo{pages}{259--269}.
\newblock


\bibitem[Brereton et~al\mbox{.}(2007)]%
        {brereton2007lessons}
\bibfield{author}{\bibinfo{person}{Pearl Brereton}, \bibinfo{person}{Barbara~A
  Kitchenham}, \bibinfo{person}{David Budgen}, \bibinfo{person}{Mark Turner},
  {and} \bibinfo{person}{Mohamed Khalil}.} \bibinfo{year}{2007}\natexlab{}.
\newblock \showarticletitle{Lessons from applying the systematic literature
  review process within the software engineering domain}.
\newblock \bibinfo{journal}{\emph{Journal of systems and software}}
  \bibinfo{volume}{80}, \bibinfo{number}{4} (\bibinfo{year}{2007}),
  \bibinfo{pages}{571--583}.
\newblock


\bibitem[Brito et~al\mbox{.}(2016)]%
        {brito2016developers}
\bibfield{author}{\bibinfo{person}{Gleison Brito}, \bibinfo{person}{Andre
  Hora}, \bibinfo{person}{Marco~Tulio Valente}, {and} \bibinfo{person}{Romain
  Robbes}.} \bibinfo{year}{2016}\natexlab{}.
\newblock \showarticletitle{Do developers deprecate apis with replacement
  messages? a large-scale analysis on java systems}. In
  \bibinfo{booktitle}{\emph{2016 IEEE 23rd International Conference on Software
  Analysis, Evolution, and Reengineering (SANER)}}, Vol.~\bibinfo{volume}{1}.
  IEEE, \bibinfo{pages}{360--369}.
\newblock


\bibitem[Cai et~al\mbox{.}(2019)]%
        {cai2019large}
\bibfield{author}{\bibinfo{person}{Haipeng Cai}, \bibinfo{person}{Ziyi Zhang},
  \bibinfo{person}{Li Li}, {and} \bibinfo{person}{Xiaoqin Fu}.}
  \bibinfo{year}{2019}\natexlab{}.
\newblock \showarticletitle{A Large-Scale Study of Application
  Incompatibilities in Android}. In \bibinfo{booktitle}{\emph{The 28th ACM
  SIGSOFT International Symposium on Software Testing and Analysis (ISSTA
  2019)}}.
\newblock


\bibitem[Chen et~al\mbox{.}(2020)]%
        {chen2020taming}
\bibfield{author}{\bibinfo{person}{Lingchao Chen}, \bibinfo{person}{Foyzul
  Hassan}, \bibinfo{person}{Xiaoyin Wang}, {and} \bibinfo{person}{Lingming
  Zhang}.} \bibinfo{year}{2020}\natexlab{}.
\newblock \showarticletitle{Taming behavioral backward incompatibilities via
  cross-project testing and analysis}. In \bibinfo{booktitle}{\emph{Proceedings
  of the ACM/IEEE 42nd International Conference on Software Engineering}}.
  \bibinfo{pages}{112--124}.
\newblock


\bibitem[Collie et~al\mbox{.}(2020)]%
        {collie2020m3}
\bibfield{author}{\bibinfo{person}{Bruce Collie}, \bibinfo{person}{Philip
  Ginsbach}, \bibinfo{person}{Jackson Woodruff}, \bibinfo{person}{Ajitha
  Rajan}, {and} \bibinfo{person}{Michael O'Boyle}.}
  \bibinfo{year}{2020}\natexlab{}.
\newblock \showarticletitle{M3: Semantic API Migrations}.
\newblock \bibinfo{journal}{\emph{arXiv preprint arXiv:2008.12118}}
  (\bibinfo{year}{2020}).
\newblock


\bibitem[Ferrante et~al\mbox{.}(1987)]%
        {ferrante1987program}
\bibfield{author}{\bibinfo{person}{Jeanne Ferrante}, \bibinfo{person}{Karl~J
  Ottenstein}, {and} \bibinfo{person}{Joe~D Warren}.}
  \bibinfo{year}{1987}\natexlab{}.
\newblock \showarticletitle{The program dependence graph and its use in
  optimization}.
\newblock \bibinfo{journal}{\emph{ACM Transactions on Programming Languages and
  Systems (TOPLAS)}} \bibinfo{volume}{9}, \bibinfo{number}{3}
  (\bibinfo{year}{1987}), \bibinfo{pages}{319--349}.
\newblock


\bibitem[Gordon et~al\mbox{.}(2015)]%
        {gordon2015information}
\bibfield{author}{\bibinfo{person}{Michael~I Gordon}, \bibinfo{person}{Deokhwan
  Kim}, \bibinfo{person}{Jeff~H Perkins}, \bibinfo{person}{Limei Gilham},
  \bibinfo{person}{Nguyen Nguyen}, {and} \bibinfo{person}{Martin~C Rinard}.}
  \bibinfo{year}{2015}\natexlab{}.
\newblock \showarticletitle{Information flow analysis of android applications
  in droidsafe.}. In \bibinfo{booktitle}{\emph{NDSS}},
  Vol.~\bibinfo{volume}{15}. \bibinfo{pages}{110}.
\newblock


\bibitem[He et~al\mbox{.}(2018)]%
        {he2018understanding}
\bibfield{author}{\bibinfo{person}{Dongjie He}, \bibinfo{person}{Lian Li},
  \bibinfo{person}{Lei Wang}, \bibinfo{person}{Hengjie Zheng},
  \bibinfo{person}{Guangwei Li}, {and} \bibinfo{person}{Jingling Xue}.}
  \bibinfo{year}{2018}\natexlab{}.
\newblock \showarticletitle{Understanding and detecting evolution-induced
  compatibility issues in Android apps}. In \bibinfo{booktitle}{\emph{2018 33rd
  IEEE/ACM International Conference on Automated Software Engineering (ASE)}}.
  IEEE, \bibinfo{pages}{167--177}.
\newblock


\bibitem[Hora et~al\mbox{.}(2015)]%
        {hora2015developers}
\bibfield{author}{\bibinfo{person}{Andr{\'e} Hora}, \bibinfo{person}{Romain
  Robbes}, \bibinfo{person}{Nicolas Anquetil}, \bibinfo{person}{Anne Etien},
  \bibinfo{person}{St{\'e}phane Ducasse}, {and} \bibinfo{person}{Marco~Tulio
  Valente}.} \bibinfo{year}{2015}\natexlab{}.
\newblock \showarticletitle{How do developers react to API evolution? The Pharo
  ecosystem case}. In \bibinfo{booktitle}{\emph{2015 IEEE International
  Conference on Software Maintenance and Evolution (ICSME)}}. IEEE,
  \bibinfo{pages}{251--260}.
\newblock


\bibitem[Huang et~al\mbox{.}(2018)]%
        {huang2018understanding}
\bibfield{author}{\bibinfo{person}{Huaxun Huang}, \bibinfo{person}{Lili Wei},
  \bibinfo{person}{Yepang Liu}, {and} \bibinfo{person}{Shing-Chi Cheung}.}
  \bibinfo{year}{2018}\natexlab{}.
\newblock \showarticletitle{Understanding and detecting callback compatibility
  issues for android applications}. In \bibinfo{booktitle}{\emph{Proceedings of
  the 33rd ACM/IEEE International Conference on Automated Software
  Engineering}}. \bibinfo{pages}{532--542}.
\newblock


\bibitem[Kang et~al\mbox{.}(2019)]%
        {kang2019semantic}
\bibfield{author}{\bibinfo{person}{Hong~Jin Kang}, \bibinfo{person}{Ferdian
  Thung}, \bibinfo{person}{Julia Lawall}, \bibinfo{person}{Gilles Muller},
  \bibinfo{person}{Lingxiao Jiang}, {and} \bibinfo{person}{David Lo}.}
  \bibinfo{year}{2019}\natexlab{}.
\newblock \showarticletitle{Semantic Patches for Java Program Transformation
  (Experience Report)}. In \bibinfo{booktitle}{\emph{33rd European Conference
  on Object-Oriented Programming (ECOOP 2019)}}. Schloss
  Dagstuhl-Leibniz-Zentrum fuer Informatik.
\newblock


\bibitem[Keele et~al\mbox{.}(2007)]%
        {keele2007guidelines}
\bibfield{author}{\bibinfo{person}{Staffs Keele} {et~al\mbox{.}}}
  \bibinfo{year}{2007}\natexlab{}.
\newblock \bibinfo{booktitle}{\emph{Guidelines for performing systematic
  literature reviews in software engineering}}.
\newblock \bibinfo{type}{{T}echnical {R}eport}.
  \bibinfo{institution}{Citeseer}.
\newblock


\bibitem[Ki et~al\mbox{.}(2019)]%
        {ki2019mimic}
\bibfield{author}{\bibinfo{person}{Taeyeon Ki}, \bibinfo{person}{Chang~Min
  Park}, \bibinfo{person}{Karthik Dantu}, \bibinfo{person}{Steven~Y Ko}, {and}
  \bibinfo{person}{Lukasz Ziarek}.} \bibinfo{year}{2019}\natexlab{}.
\newblock \showarticletitle{Mimic: UI compatibility testing system for Android
  apps}. In \bibinfo{booktitle}{\emph{2019 IEEE/ACM 41st International
  Conference on Software Engineering (ICSE)}}. IEEE, \bibinfo{pages}{246--256}.
\newblock


\bibitem[Kong et~al\mbox{.}(2018)]%
        {kong2018automated}
\bibfield{author}{\bibinfo{person}{Pingfan Kong}, \bibinfo{person}{Li Li},
  \bibinfo{person}{Jun Gao}, \bibinfo{person}{Kui Liu},
  \bibinfo{person}{Tegawend{\'e}~F Bissyand{\'e}}, {and}
  \bibinfo{person}{Jacques Klein}.} \bibinfo{year}{2018}\natexlab{}.
\newblock \showarticletitle{Automated Testing of Android Apps: A Systematic
  Literature Review}.
\newblock \bibinfo{journal}{\emph{IEEE Transactions on Reliability}}
  (\bibinfo{year}{2018}).
\newblock


\bibitem[Lawall and Muller(2018)]%
        {lawall2018coccinelle}
\bibfield{author}{\bibinfo{person}{Julia Lawall} {and} \bibinfo{person}{Gilles
  Muller}.} \bibinfo{year}{2018}\natexlab{}.
\newblock \showarticletitle{Coccinelle: 10 years of automated evolution in the
  Linux kernel}. In \bibinfo{booktitle}{\emph{2018 $\{$USENIX$\}$ Annual
  Technical Conference ($\{$USENIX$\}$\{$ATC$\} 18)}}.
  \bibinfo{pages}{601--614}.
\newblock


\bibitem[Li et~al\mbox{.}(2015)]%
        {li2015iccta}
\bibfield{author}{\bibinfo{person}{Li Li}, \bibinfo{person}{Alexandre Bartel},
  \bibinfo{person}{Tegawend{\'e}~F Bissyand{\'e}}, \bibinfo{person}{Jacques
  Klein}, \bibinfo{person}{Yves Le~Traon}, \bibinfo{person}{Steven Arzt},
  \bibinfo{person}{Siegfried Rasthofer}, \bibinfo{person}{Eric Bodden},
  \bibinfo{person}{Damien Octeau}, {and} \bibinfo{person}{Patrick McDaniel}.}
  \bibinfo{year}{2015}\natexlab{}.
\newblock \showarticletitle{Iccta: Detecting inter-component privacy leaks in
  android apps}. In \bibinfo{booktitle}{\emph{2015 IEEE/ACM 37th IEEE
  International Conference on Software Engineering}}, Vol.~\bibinfo{volume}{1}.
  IEEE, \bibinfo{pages}{280--291}.
\newblock


\bibitem[Li et~al\mbox{.}(2018a)]%
        {li2018moonlightbox}
\bibfield{author}{\bibinfo{person}{Li Li}, \bibinfo{person}{Tegawend{\'e}~F
  Bissyand{\'e}}, {and} \bibinfo{person}{Jacques Klein}.}
  \bibinfo{year}{2018}\natexlab{a}.
\newblock \showarticletitle{MoonlightBox: Mining Android API Histories for
  Uncovering Release-time Inconsistencies}. In \bibinfo{booktitle}{\emph{The
  29th IEEE International Symposium on Software Reliability Engineering (ISSRE
  2018)}}.
\newblock


\bibitem[Li et~al\mbox{.}(2016a)]%
        {li2016accessing}
\bibfield{author}{\bibinfo{person}{Li Li}, \bibinfo{person}{Tegawend{\'e}~F
  Bissyand{\'e}}, \bibinfo{person}{Yves Le~Traon}, {and}
  \bibinfo{person}{Jacques Klein}.} \bibinfo{year}{2016}\natexlab{a}.
\newblock \showarticletitle{Accessing Inaccessible Android APIs: An Empirical
  Study}. In \bibinfo{booktitle}{\emph{The 32nd International Conference on
  Software Maintenance and Evolution (ICSME 2016)}}.
\newblock


\bibitem[Li et~al\mbox{.}(2016b)]%
        {li2016droidra}
\bibfield{author}{\bibinfo{person}{Li Li}, \bibinfo{person}{Tegawend{\'e}~F
  Bissyand{\'e}}, \bibinfo{person}{Damien Octeau}, {and}
  \bibinfo{person}{Jacques Klein}.} \bibinfo{year}{2016}\natexlab{b}.
\newblock \showarticletitle{DroidRA: Taming Reflection to Support Whole-Program
  Analysis of Android Apps}. In \bibinfo{booktitle}{\emph{The 2016
  International Symposium on Software Testing and Analysis (ISSTA 2016)}}.
\newblock


\bibitem[Li et~al\mbox{.}(2017)]%
        {li2017static}
\bibfield{author}{\bibinfo{person}{Li Li}, \bibinfo{person}{Tegawend{\'e}~F
  Bissyand{\'e}}, \bibinfo{person}{Mike Papadakis}, \bibinfo{person}{Siegfried
  Rasthofer}, \bibinfo{person}{Alexandre Bartel}, \bibinfo{person}{Damien
  Octeau}, \bibinfo{person}{Jacques Klein}, {and} \bibinfo{person}{Yves
  Le~Traon}.} \bibinfo{year}{2017}\natexlab{}.
\newblock \showarticletitle{Static Analysis of Android Apps: A Systematic
  Literature Review}.
\newblock \bibinfo{journal}{\emph{Information and Software Technology}}
  (\bibinfo{year}{2017}).
\newblock


\bibitem[Li et~al\mbox{.}(2018b)]%
        {li2018cid}
\bibfield{author}{\bibinfo{person}{Li Li}, \bibinfo{person}{Tegawend{\'e}~F
  Bissyand{\'e}}, \bibinfo{person}{Haoyu Wang}, {and} \bibinfo{person}{Jacques
  Klein}.} \bibinfo{year}{2018}\natexlab{b}.
\newblock \showarticletitle{Cid: Automating the detection of api-related
  compatibility issues in android apps}. In
  \bibinfo{booktitle}{\emph{Proceedings of the 27th ACM SIGSOFT International
  Symposium on Software Testing and Analysis}}. \bibinfo{pages}{153--163}.
\newblock


\bibitem[Li et~al\mbox{.}(2018c)]%
        {li2018characterising}
\bibfield{author}{\bibinfo{person}{Li Li}, \bibinfo{person}{Jun Gao},
  \bibinfo{person}{Tegawend{\'e}~F Bissyand{\'e}}, \bibinfo{person}{Lei Ma},
  \bibinfo{person}{Xin Xia}, {and} \bibinfo{person}{Jacques Klein}.}
  \bibinfo{year}{2018}\natexlab{c}.
\newblock \showarticletitle{Characterising deprecated android apis}. In
  \bibinfo{booktitle}{\emph{Proceedings of the 15th International Conference on
  Mining Software Repositories}}. \bibinfo{pages}{254--264}.
\newblock


\bibitem[Li et~al\mbox{.}(2020)]%
        {li2020cda}
\bibfield{author}{\bibinfo{person}{Li Li}, \bibinfo{person}{Jun Gao},
  \bibinfo{person}{Tegawend{\'e}~F Bissyand{\'e}}, \bibinfo{person}{Lei Ma},
  \bibinfo{person}{Xin Xia}, {and} \bibinfo{person}{Jacques Klein}.}
  \bibinfo{year}{2020}\natexlab{}.
\newblock \showarticletitle{Cda: Characterising deprecated android apis}.
\newblock \bibinfo{journal}{\emph{Empirical Software Engineering}}
  (\bibinfo{year}{2020}), \bibinfo{pages}{1--41}.
\newblock


\bibitem[Liu et~al\mbox{.}(2021a)]%
        {liu2021identifying}
\bibfield{author}{\bibinfo{person}{Pei Liu}, \bibinfo{person}{Li Li},
  \bibinfo{person}{Yichun Yan}, \bibinfo{person}{Mattia Fazzini}, {and}
  \bibinfo{person}{John Grundy}.} \bibinfo{year}{2021}\natexlab{a}.
\newblock \showarticletitle{Identifying and Characterizing Silently-Evolved
  Methods in the Android API}. In \bibinfo{booktitle}{\emph{The 43rd ACM/IEEE
  International Conference on Software Engineering, SEIP Track (ICSE-SEIP
  2021)}}.
\newblock


\bibitem[Liu et~al\mbox{.}(2021b)]%
        {liu2021deep}
\bibfield{author}{\bibinfo{person}{Yue Liu}, \bibinfo{person}{Chakkrit
  Tantithamthavorn}, \bibinfo{person}{Li Li}, {and} \bibinfo{person}{Yepang
  Liu}.} \bibinfo{year}{2021}\natexlab{b}.
\newblock \showarticletitle{Deep learning for android malware defenses: a
  systematic literature review}.
\newblock \bibinfo{journal}{\emph{arXiv preprint arXiv:2103.05292}}
  (\bibinfo{year}{2021}).
\newblock


\bibitem[Mahmud et~al\mbox{.}(2021)]%
        {mahmud2021android}
\bibfield{author}{\bibinfo{person}{Tarek Mahmud}, \bibinfo{person}{Meiru Che},
  {and} \bibinfo{person}{Guowei Yang}.} \bibinfo{year}{2021}\natexlab{}.
\newblock \showarticletitle{Android Compatibility Issue Detection Using API
  Differences}. In \bibinfo{booktitle}{\emph{2021 IEEE International Conference
  on Software Analysis, Evolution and Reengineering (SANER)}}. IEEE,
  \bibinfo{pages}{480--490}.
\newblock


\bibitem[Mobilio et~al\mbox{.}(2020)]%
        {mobilio2020filo}
\bibfield{author}{\bibinfo{person}{Marco Mobilio}, \bibinfo{person}{Oliviero
  Riganelli}, \bibinfo{person}{Daniela Micucci}, {and}
  \bibinfo{person}{Leonardo Mariani}.} \bibinfo{year}{2020}\natexlab{}.
\newblock \showarticletitle{FILO: FIx-LOcus localization for backward
  incompatibilities caused by Android framework upgrades}. In
  \bibinfo{booktitle}{\emph{Proceedings of the 35th IEEE/ACM International
  Conference on Automated Software Engineering}}. \bibinfo{pages}{1292--1296}.
\newblock


\bibitem[Nielebock et~al\mbox{.}(2021)]%
        {nielebock2021androidcompass}
\bibfield{author}{\bibinfo{person}{Sebastian Nielebock}, \bibinfo{person}{Paul
  Blockhaus}, \bibinfo{person}{Jacob Kr{\"u}ger}, {and} \bibinfo{person}{Frank
  Ortmeier}.} \bibinfo{year}{2021}\natexlab{}.
\newblock \showarticletitle{AndroidCompass: A Dataset of Android Compatibility
  Checks in Code Repositories}.
\newblock \bibinfo{journal}{\emph{arXiv preprint arXiv:2103.09620}}
  (\bibinfo{year}{2021}).
\newblock


\bibitem[Qiu et~al\mbox{.}(2018)]%
        {qiu2018analyzing}
\bibfield{author}{\bibinfo{person}{Lina Qiu}, \bibinfo{person}{Yingying Wang},
  {and} \bibinfo{person}{Julia Rubin}.} \bibinfo{year}{2018}\natexlab{}.
\newblock \showarticletitle{Analyzing the analyzers: Flowdroid/iccta,
  amandroid, and droidsafe}. In \bibinfo{booktitle}{\emph{Proceedings of the
  27th ACM SIGSOFT International Symposium on Software Testing and Analysis}}.
  \bibinfo{pages}{176--186}.
\newblock


\bibitem[Robbes et~al\mbox{.}(2012)]%
        {robbes2012developers}
\bibfield{author}{\bibinfo{person}{Romain Robbes}, \bibinfo{person}{Mircea
  Lungu}, {and} \bibinfo{person}{David R{\"o}thlisberger}.}
  \bibinfo{year}{2012}\natexlab{}.
\newblock \showarticletitle{How do developers react to API deprecation? The
  case of a Smalltalk ecosystem}. In \bibinfo{booktitle}{\emph{Proceedings of
  the ACM SIGSOFT 20th International Symposium on the Foundations of Software
  Engineering}}. \bibinfo{pages}{1--11}.
\newblock


\bibitem[Rolim et~al\mbox{.}(2017)]%
        {rolim2017learning}
\bibfield{author}{\bibinfo{person}{Reudismam Rolim}, \bibinfo{person}{Gustavo
  Soares}, \bibinfo{person}{Loris D'Antoni}, \bibinfo{person}{Oleksandr
  Polozov}, \bibinfo{person}{Sumit Gulwani}, \bibinfo{person}{Rohit Gheyi},
  \bibinfo{person}{Ryo Suzuki}, {and} \bibinfo{person}{Bj{\"o}rn Hartmann}.}
  \bibinfo{year}{2017}\natexlab{}.
\newblock \showarticletitle{Learning syntactic program transformations from
  examples}. In \bibinfo{booktitle}{\emph{2017 IEEE/ACM 39th International
  Conference on Software Engineering (ICSE)}}. IEEE, \bibinfo{pages}{404--415}.
\newblock


\bibitem[Sawant et~al\mbox{.}(2018)]%
        {sawant2018reaction}
\bibfield{author}{\bibinfo{person}{Anand~Ashok Sawant}, \bibinfo{person}{Romain
  Robbes}, {and} \bibinfo{person}{Alberto Bacchelli}.}
  \bibinfo{year}{2018}\natexlab{}.
\newblock \showarticletitle{On the reaction to deprecation of clients of 4+ 1
  popular Java APIs and the JDK}.
\newblock \bibinfo{journal}{\emph{Empirical Software Engineering}}
  \bibinfo{volume}{23}, \bibinfo{number}{4} (\bibinfo{year}{2018}),
  \bibinfo{pages}{2158--2197}.
\newblock


\bibitem[Scalabrino et~al\mbox{.}(2019)]%
        {scalabrino2019data}
\bibfield{author}{\bibinfo{person}{Simone Scalabrino},
  \bibinfo{person}{Gabriele Bavota}, \bibinfo{person}{Mario
  Linares-V{\'a}squez}, \bibinfo{person}{Michele Lanza}, {and}
  \bibinfo{person}{Rocco Oliveto}.} \bibinfo{year}{2019}\natexlab{}.
\newblock \showarticletitle{Data-driven solutions to detect api compatibility
  issues in android: an empirical study}. In \bibinfo{booktitle}{\emph{2019
  IEEE/ACM 16th International Conference on Mining Software Repositories
  (MSR)}}. IEEE, \bibinfo{pages}{288--298}.
\newblock


\bibitem[Scalabrino et~al\mbox{.}(2020)]%
        {scalabrino2020api}
\bibfield{author}{\bibinfo{person}{Simone Scalabrino},
  \bibinfo{person}{Gabriele Bavota}, \bibinfo{person}{Mario
  Linares-V{\'a}squez}, \bibinfo{person}{Valentina Piantadosi},
  \bibinfo{person}{Michele Lanza}, {and} \bibinfo{person}{Rocco Oliveto}.}
  \bibinfo{year}{2020}\natexlab{}.
\newblock \showarticletitle{API compatibility issues in Android: Causes and
  effectiveness of data-driven detection techniques}.
\newblock \bibinfo{journal}{\emph{Empirical Software Engineering}}
  \bibinfo{volume}{25}, \bibinfo{number}{6} (\bibinfo{year}{2020}),
  \bibinfo{pages}{5006--5046}.
\newblock


\bibitem[Shamsujjoha et~al\mbox{.}(2021)]%
        {shamsujjoha2021developing}
\bibfield{author}{\bibinfo{person}{Md. Shamsujjoha}, \bibinfo{person}{John
  Grundy}, \bibinfo{person}{Li Li}, \bibinfo{person}{Hourieh Khalajzadeh},
  {and} \bibinfo{person}{Qinghua Lu}.} \bibinfo{year}{2021}\natexlab{}.
\newblock \showarticletitle{Developing Mobile Applications via Model Driven
  Development: A Systematic Literature Review}.
\newblock \bibinfo{journal}{\emph{Information and Software Technology (IST)}}
  (\bibinfo{year}{2021}).
\newblock


\bibitem[Su et~al\mbox{.}(2021)]%
        {su2021benchmarking}
\bibfield{author}{\bibinfo{person}{Ting Su}, \bibinfo{person}{Jue Wang}, {and}
  \bibinfo{person}{Zhendong Su}.} \bibinfo{year}{2021}\natexlab{}.
\newblock \showarticletitle{Benchmarking automated GUI testing for Android
  against real-world bugs}. In \bibinfo{booktitle}{\emph{Proceedings of the
  29th ACM Joint Meeting on European Software Engineering Conference and
  Symposium on the Foundations of Software Engineering}}.
  \bibinfo{pages}{119--130}.
\newblock


\bibitem[Sun et~al\mbox{.}(2020)]%
        {sun2020taming}
\bibfield{author}{\bibinfo{person}{Xiaoyu Sun}, \bibinfo{person}{Li Li},
  \bibinfo{person}{Tegawend{\'e}~F Bissyand{\'e}}, \bibinfo{person}{Jacques
  Klein}, \bibinfo{person}{Damien Octeau}, {and} \bibinfo{person}{John
  Grundy}.} \bibinfo{year}{2020}\natexlab{}.
\newblock \showarticletitle{Taming Reflection: An Essential Step Towards
  Whole-Program Analysis of Android Apps}.
\newblock \bibinfo{journal}{\emph{ACM Transactions on Software Engineering and
  Methodology (TOSEM)}} (\bibinfo{year}{2020}).
\newblock


\bibitem[Vall{\'e}e-Rai et~al\mbox{.}(2010)]%
        {vallee2010soot}
\bibfield{author}{\bibinfo{person}{Raja Vall{\'e}e-Rai}, \bibinfo{person}{Phong
  Co}, \bibinfo{person}{Etienne Gagnon}, \bibinfo{person}{Laurie Hendren},
  \bibinfo{person}{Patrick Lam}, {and} \bibinfo{person}{Vijay Sundaresan}.}
  \bibinfo{year}{2010}\natexlab{}.
\newblock \showarticletitle{Soot: A Java bytecode optimization framework}.
\newblock In \bibinfo{booktitle}{\emph{CASCON First Decade High Impact
  Papers}}. \bibinfo{pages}{214--224}.
\newblock


\bibitem[Wang et~al\mbox{.}(2019)]%
        {wang2019characterizing}
\bibfield{author}{\bibinfo{person}{Haoyu Wang}, \bibinfo{person}{Hongxuan Liu},
  \bibinfo{person}{Xusheng Xiao}, \bibinfo{person}{Guozhu Meng}, {and}
  \bibinfo{person}{Yao Guo}.} \bibinfo{year}{2019}\natexlab{}.
\newblock \showarticletitle{Characterizing Android app signing issues}. In
  \bibinfo{booktitle}{\emph{2019 34th IEEE/ACM International Conference on
  Automated Software Engineering (ASE)}}. IEEE, \bibinfo{pages}{280--292}.
\newblock


\bibitem[Wei et~al\mbox{.}(2014)]%
        {wei2014amandroid}
\bibfield{author}{\bibinfo{person}{Fengguo Wei}, \bibinfo{person}{Sankardas
  Roy}, {and} \bibinfo{person}{Xinming Ou}.} \bibinfo{year}{2014}\natexlab{}.
\newblock \showarticletitle{Amandroid: A precise and general inter-component
  data flow analysis framework for security vetting of android apps}. In
  \bibinfo{booktitle}{\emph{Proceedings of the 2014 ACM SIGSAC conference on
  computer and communications security}}. \bibinfo{pages}{1329--1341}.
\newblock


\bibitem[Wei et~al\mbox{.}(2018b)]%
        {wei2018amandroid}
\bibfield{author}{\bibinfo{person}{Fengguo Wei}, \bibinfo{person}{Sankardas
  Roy}, {and} \bibinfo{person}{Xinming Ou}.} \bibinfo{year}{2018}\natexlab{b}.
\newblock \showarticletitle{Amandroid: A precise and general inter-component
  data flow analysis framework for security vetting of android apps}.
\newblock \bibinfo{journal}{\emph{ACM Transactions on Privacy and Security
  (TOPS)}} \bibinfo{volume}{21}, \bibinfo{number}{3} (\bibinfo{year}{2018}),
  \bibinfo{pages}{1--32}.
\newblock


\bibitem[Wei et~al\mbox{.}(2016)]%
        {wei2016taming}
\bibfield{author}{\bibinfo{person}{Lili Wei}, \bibinfo{person}{Yepang Liu},
  {and} \bibinfo{person}{Shing-Chi Cheung}.} \bibinfo{year}{2016}\natexlab{}.
\newblock \showarticletitle{Taming Android fragmentation: Characterizing and
  detecting compatibility issues for Android apps}. In
  \bibinfo{booktitle}{\emph{Proceedings of the 31st IEEE/ACM International
  Conference on Automated Software Engineering}}. \bibinfo{pages}{226--237}.
\newblock


\bibitem[Wei et~al\mbox{.}(2019)]%
        {wei2019pivot}
\bibfield{author}{\bibinfo{person}{Lili Wei}, \bibinfo{person}{Yepang Liu},
  {and} \bibinfo{person}{Shing-Chi Cheung}.} \bibinfo{year}{2019}\natexlab{}.
\newblock \showarticletitle{Pivot: learning api-device correlations to
  facilitate android compatibility issue detection}. In
  \bibinfo{booktitle}{\emph{2019 IEEE/ACM 41st International Conference on
  Software Engineering (ICSE)}}. IEEE, \bibinfo{pages}{878--888}.
\newblock


\bibitem[Wei et~al\mbox{.}(2018a)]%
        {wei2018understanding}
\bibfield{author}{\bibinfo{person}{Lili Wei}, \bibinfo{person}{Yepang Liu},
  \bibinfo{person}{Shing-Chi Cheung}, \bibinfo{person}{Huaxun Huang},
  \bibinfo{person}{Xuan Lu}, {and} \bibinfo{person}{Xuanzhe Liu}.}
  \bibinfo{year}{2018}\natexlab{a}.
\newblock \showarticletitle{Understanding and detecting fragmentation-induced
  compatibility issues for android apps}.
\newblock \bibinfo{journal}{\emph{IEEE Transactions on Software Engineering}}
  \bibinfo{volume}{46}, \bibinfo{number}{11} (\bibinfo{year}{2018}),
  \bibinfo{pages}{1176--1199}.
\newblock


\bibitem[Xia et~al\mbox{.}(2020)]%
        {xia2020android}
\bibfield{author}{\bibinfo{person}{Hao Xia}, \bibinfo{person}{Yuan Zhang},
  \bibinfo{person}{Yingtian Zhou}, \bibinfo{person}{Xiaoting Chen},
  \bibinfo{person}{Yang Wang}, \bibinfo{person}{Xiangyu Zhang},
  \bibinfo{person}{Shuaishuai Cui}, \bibinfo{person}{Geng Hong},
  \bibinfo{person}{Xiaohan Zhang}, \bibinfo{person}{Min Yang}, {et~al\mbox{.}}}
  \bibinfo{year}{2020}\natexlab{}.
\newblock \showarticletitle{How Android developers handle evolution-induced API
  compatibility issues: a large-scale study}. In \bibinfo{booktitle}{\emph{2020
  IEEE/ACM 42nd International Conference on Software Engineering (ICSE)}}.
  IEEE, \bibinfo{pages}{886--898}.
\newblock


\bibitem[Yang et~al\mbox{.}(2015)]%
        {yang2015static}
\bibfield{author}{\bibinfo{person}{Shengqian Yang}, \bibinfo{person}{Dacong
  Yan}, \bibinfo{person}{Haowei Wu}, \bibinfo{person}{Yan Wang}, {and}
  \bibinfo{person}{Atanas Rountev}.} \bibinfo{year}{2015}\natexlab{}.
\newblock \showarticletitle{Static control-flow analysis of user-driven
  callbacks in Android applications}. In \bibinfo{booktitle}{\emph{2015
  IEEE/ACM 37th IEEE International Conference on Software Engineering}},
  Vol.~\bibinfo{volume}{1}. IEEE, \bibinfo{pages}{89--99}.
\newblock


\bibitem[Zhan et~al\mbox{.}(2021)]%
        {zhan2021research}
\bibfield{author}{\bibinfo{person}{Xian Zhan}, \bibinfo{person}{Tianming Liu},
  \bibinfo{person}{Lingling Fan}, \bibinfo{person}{Li Li}, \bibinfo{person}{Sen
  Chen}, \bibinfo{person}{Xiapu Luo}, {and} \bibinfo{person}{Yang Liu}.}
  \bibinfo{year}{2021}\natexlab{}.
\newblock \showarticletitle{Research on Third-Party Libraries in Android Apps:
  A Taxonomy and Systematic Literature Review}.
\newblock \bibinfo{journal}{\emph{IEEE Transactions on Software Engineering}}
  (\bibinfo{year}{2021}).
\newblock


\bibitem[Zhao et~al\mbox{.}(2022)]%
        {zhao2022towards}
\bibfield{author}{\bibinfo{person}{Yanjie Zhao}, \bibinfo{person}{Li Li},
  \bibinfo{person}{Kui Liu}, {and} \bibinfo{person}{John Grundy}.}
  \bibinfo{year}{2022}\natexlab{}.
\newblock \showarticletitle{Towards Automatically Repairing Compatibility
  Issues in Published Android Apps}. In \bibinfo{booktitle}{\emph{The 44th
  International Conference on Software Engineering (ICSE 2022)}}.
\newblock


\bibitem[Zhou and Walker(2016)]%
        {zhou2016api}
\bibfield{author}{\bibinfo{person}{Jing Zhou} {and} \bibinfo{person}{Robert~J
  Walker}.} \bibinfo{year}{2016}\natexlab{}.
\newblock \showarticletitle{API deprecation: a retrospective analysis and
  detection method for code examples on the web}. In
  \bibinfo{booktitle}{\emph{Proceedings of the 2016 24th ACM SIGSOFT
  International Symposium on Foundations of Software Engineering}}.
  \bibinfo{pages}{266--277}.
\newblock


\end{thebibliography}

\end{document}